\documentclass[natphys,reprint,twocolumn,superscriptaddress,floatfix,nofootinbib,longbibliography]{revtex4-2}
\usepackage{natbib}
\usepackage{graphicx,amsmath,amsfonts,amssymb,amsthm,xr,braket}
\usepackage{framed}
\usepackage{epsfig,amsmath,amssymb,color,dsfont,upgreek,physics}
\usepackage{mathrsfs}
\usepackage{mathtools}
\usepackage{bbold}
\usepackage{comment}
\usepackage{float}

\usepackage[bookmarks=true,colorlinks,linkcolor=OrangeRed,urlcolor=NavyBlue,citecolor=RoyalBlue]{hyperref}
\usepackage[dvipsnames]{xcolor}

\definecolor{mygold}{rgb}{0.93,0.69,0.13}

\definecolor{mypurple}{rgb}{0.49,0.18,0.56}

\definecolor{mygreen}{rgb}{0,0.5,0}

\definecolor{mygreen}{rgb}{0,0.5,0}
\definecolor{myred}{rgb}{0.7,0,0}
\definecolor{myblue}{rgb}{0,0,0.5}

\usepackage{bm}	
\renewcommand{\vec}[1]{\bm{#1}}
\renewcommand{\ij}{{\langle \vec{i}, \vec{j} \rangle}}

\renewcommand{\a}{\hat{a}}
\newcommand{\ad}{\hat{a}^\dagger}
\newcommand{\hc}{\rm H.c. }

\begin{document}
\title{Cold-atom quantum simulators of gauge theories}
\author{Jad C.~Halimeh}
\email{jad.halimeh@physik.lmu.de}
\affiliation{Munich Center for Quantum Science and Technology (MCQST), Schellingstra\ss e 4, D-80799 M\"unchen, Germany}
\affiliation{Department of Physics and Arnold Sommerfeld Center for Theoretical Physics (ASC), Ludwig-Maximilians-Universit\"at M\"unchen, Theresienstra\ss e 37, D-80333 M\"unchen, Germany}
\author{Monika Aidelsburger}
\affiliation{Max-Planck-Institut f\"ur Quantenoptik, 85748 Garching, Germany}
\affiliation{Faculty of Physics, Ludwig-Maximilians-Universit\"at M\"unchen, Schellingstra\ss e 4, D-80799 Munich, Germany}
\affiliation{Munich Center for Quantum Science and Technology (MCQST), Schellingstra\ss e 4, D-80799 M\"unchen, Germany}
\author{Fabian Grusdt}
\affiliation{Munich Center for Quantum Science and Technology (MCQST), Schellingstra\ss e 4, D-80799 M\"unchen, Germany}
\affiliation{Department of Physics and Arnold Sommerfeld Center for Theoretical Physics (ASC), Ludwig-Maximilians-Universit\"at M\"unchen, Theresienstra\ss e 37, D-80333 M\"unchen, Germany}
\author{Philipp Hauke}
\affiliation{Pitaevskii BEC Center and Department of Physics, University of Trento, Via Sommarive 14, I-38123 Trento, Italy}
\affiliation{INFN-TIFPA, Trento Institute for Fundamental Physics and Applications, Trento, Italy}
\author{Bing Yang}
\email{yangbing@sustech.edu.cn}
\affiliation{Department of Physics, Southern University of Science and Technology, Shenzhen 518055, China}

\begin{abstract}
Gauge theories represent a fundamental framework underlying modern physics, constituting the basis of the Standard Model and also providing useful descriptions of various phenomena in condensed matter. Realizing gauge theories on accessible and tunable tabletop quantum devices offers the possibility to study their dynamics from first principles time evolution and to probe their exotic physics, including that generated by deviations from gauge invariance, which is not possible, e.g., in dedicated particle colliders. Not only do cold-atom quantum simulators hold the potential to provide new insights into outstanding high-energy and nuclear-physics questions, they also provide a versatile tool for the exploration of topological phases and ergodicity-breaking mechanisms relevant to low-energy many-body physics. In recent years, cold-atom quantum simulators have demonstrated impressive progress in the large-scale implementation of $1+1$D Abelian gauge theories. In this Review, we chronicle the progress of cold-atom quantum simulators of gauge theories, highlighting the crucial advancements achieved along the way in order to reliably stabilize gauge invariance and go from building blocks to large-scale realizations where \textit{bona fide} gauge-theory phenomena can be probed. We also provide a brief outlook on where this field is heading, and what is required experimentally and theoretically to bring the technology to the next level by surveying various concrete proposals for advancing these setups to higher spatial dimensions, higher-spin representations of the gauge field, and non-Abelian gauge groups.
\end{abstract}
\date{\today}
\maketitle
\tableofcontents

\section{Introduction}
Gauge theories are a fundamental framework of modern physics that encode the laws of nature through intrinsic local relations between the distribution of matter and gauge fields~\cite{Weinberg_book}. These local relations arise directly from \textit{gauge symmetry}, which is the principal property of a gauge theory. Gauge theories are the staple of the Standard Model of particle physics, describing interactions between elementary particles as mediated through gauge bosons.

The properties of gauge theories are experimentally probed at dedicated particle colliders such as CERN's Large Hadron Collider (LHC) and Brookhaven National Laboratory's Relativistic Heavy-Ion Collider (RHIC), which provide us with a plethora of data into the inner workings of nature~\cite{Ellis_book}. The processes probed in such facilities naturally involve strongly interacting matter undergoing nonperturbative far-from-equilibrium dynamics, which is notoriously hard to treat theoretically. A prominent example is the interaction cross section in heavy-ion collisions, the physics of which is not completely understood, but which is necessary for a better understanding of hadronization~\cite{Berges_review}. Currently, various phenomenological and numerical models are used to interpret output from particle colliders and probe this cross section on classical computers, but they are fundamentally limited in their scope as they lack the capability to adequately handle the buildup of entanglement generated during the dynamics. This motivates employing \textit{quantum simulators of gauge theories}. These specialized quantum computers implement a desired model Hamiltonian in a physical and highly-controllable device whose constituents obey the laws of quantum mechanics, and thus naturally incorporate entanglement through the wave function. Such machines make it possible to access and probe time evolution from first-principles, and to even provide \textit{temporal snapshots} of the underlying nonperturbative dynamics~\cite{Wiese2013,Dalmonte_2016jk,Zohar2015,banuls2020simulating,aidelsburger_cold_2022}.

Another major front where quantum simulators of gauge theories are of crucial importance is the equilibration of isolated quantum many-body systems \cite{Eisert2015}. Whereas generic interacting many-body models are expected to thermalize according to the eigenstate thermalization hypothesis (ETH) \cite{Deutsch1991,Srednicki1994,Deutsch_review,Rigol_review}, there are many interacting systems that violate the ETH and avoid thermalization up to practically all accessible times \cite{Alet_review,Abanin_review}. Thermalization or its avoidance in many-body models is investigated using advanced numerically controlled methods such as tensor networks \cite{Orus2019}. These methods are mostly limited to (quasi-)one spatial dimension and short evolution times due to, again, entanglement buildup, although these methods have also seen important progress throughout the last years \cite{Dalmonte_review}. With their expected quantum advantage, quantum simulators of gauge theories promise to provide an ideal platform to probe the equilibration of many-body models in higher spatial dimensions and at long evolution times. The ability to quantum-simulate gauge theories is also particularly attractive in this regard, as gauge symmetry affords a very powerful knob that has been shown to give rise to exotic nonthermal dynamics \cite{Surace2020,Smith2017,Brenes2018}.

Gauge theories also emerge as effective descriptions in condensed matter, particularly in strongly correlated systems with fractionalized excitations \cite{Balents_NatureReview,Savary2016} and in finite-temperature superconductors \cite{Wen1996theory,Senthil2000Z2,Auerbach_book,Sachdev2019topological}. This serves as further motivation to realize large-scale quantum simulators of gauge theories in order to probe the rich physics of such models, especially when it is often prohibitively difficult to do so using analytic or numerical means.

\begin{figure}
	\centering
	\includegraphics[width=\columnwidth]{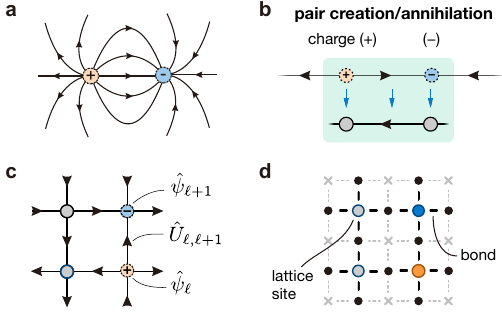}
	\caption{\textbf{Schematic illustration of Gauss's law and quantum simulation of LGTs.} 
    (a) Electric field distribution of an electric dipole in the continuum.
    (b) Illustration of Gauss's law on a link. Top: one positive (+) and one negative (-) charge on a lattice that are connected by electric field lines defined on the link (arrows). Bottom: distribution after pair annihilation, i.e., a lattice without charges.
    (c) Lattice gauge theory in 2D, where charges on the vertices are defined by the operators $\hat{\psi}_{\ell}$ and the field, which lives on the links of the 2D lattices is described by the operator $\hat{U}_{\ell,\ell+1}$.
    (d) Implementation on a 2D optical square lattice with cold atoms. The crosses mark sites that are blocked by local energy offsets, where atoms cannot tunnel onto (dashed lines). The solid lines mark tunneling bonds. They connect matter sites (gray, blue, orange circles) with link sites (black circles). This figure was adapted from Ref.~\cite{surace2023abinitio}.
    }
	\label{fig:lgt_schematic}
\end{figure}

There are various platforms of quantum simulators on which gauge theories have been realized. While significant progress has been made also on experiments in trapped ions (see, e.g., \cite{Martinez2016,Kokail2019,Mueller2023}) and superconducting qubits (see, e.g., \cite{Klco2018,Wang2021,Mildenberger2022,Huffman2022}), in this Review we will focus on \textit{cold-atom} quantum simulators~\cite{Gross2017}, which have recently emerged as viable large-scale platforms to probe the physics underlying gauge theories. This complements recent reviews on related aspects, in particular digital platforms and grand challenges for gauge-theory quantum simulation \cite{Zohar_NewReview,Bauer_review,dimeglio2023quantum}, and tensor networks \cite{Dalmonte_review}. 

The rest of this review is organized as follows. In Sec.~\ref{sec:LGT}, we briefly review lattice gauge theories (LGTs) and introduce the notation to be followed in this review. In Sec.~\ref{sec:QSCA}, we review recent advances in quantum-simulation techniques for LGTs, before going over first building block of LGT quantum simulators in Sec.~\ref{sec:BuildingBlocks}. We then cover the current state of the art in cold-atom quantum simulators of LGTs in Sec.~\ref{sec:LargeScaleCAQS}, before concluding and providing perspective in Sec.~\ref{sec:perspective}.

\section{Lattice gauge theories}\label{sec:LGT}
Due to the computational intractability of evaluating infinite-dimensional gauge-theory path integrals on a continuous spacetime, lattice gauge theories (LGTs) have been proposed as a useful tool. They involve discretizing spacetime to make the path integrals finite-dimensional, and thus amenable for various numerical treatments \cite{Rothe_book}. Indeed, lattice quantum chromodynamics (QCD) \cite{Wilson1974} is a very powerful nonperturbative approach that has led to significant progress in calculating, through Monte Carlo methods, low-energy spectra and hadron masses \cite{Ratti_review}. However, when it comes to finite baryon densities or dynamics, the sign problem renders Monte Carlo methods ineffective. Given also the limitations of tensor networks when it comes to longer evolution times and higher spatial dimensions, quantum simulators become an attractive venue to explore. Indeed, LGTs are well-suited for realization in quantum simulators, due to the discretized nature of the latter. To effectively simulate LGTs, it is crucial to create a scalable artificial quantum system with controlled interactions between constituent elements, which will be discussed in this review.

\twocolumngrid
\definecolor{shadecolor}{rgb}{0.8,0.8,0.8}
\begin{shaded}
\noindent{\bf Box 1 $|$ Quantum link formulation of lattice QED}
\end{shaded}
\vspace{-9mm}
\definecolor{shadecolor}{rgb}{0.9,0.9,0.9}
\begin{shaded}
\noindent
Lattice QED in $1+1$D includes fermionic matter degrees of $\hat{\psi}_\ell$ on sites $\ell$, and gauge and electric-field operators $\hat{U}_{\ell,\ell+1}$ and $\hat{E}_{\ell,\ell+1}$, respectively, on links in between these sites with an infinite-dimensional Hilbert space locally; see Fig.~\ref{fig:lgt_schematic}c. The link operators satisfy the commutation relations
\begin{subequations}\label{eq:comm}
    \begin{align}\label{eq:commA}
        \big[\hat{U}_{\ell,\ell+1},\hat{U}_{r,r+1}^\dagger\big]&=0,\\\label{eq:commB}
\big[\hat{E}_{\ell,\ell+1},\hat{U}_{r,r+1}\big]&=g\delta_{\ell,r}\hat{U}_{\ell,\ell+1},
\end{align}
\end{subequations}
where $g$ is the gauge-coupling strength.

For numerical and experimental feasibility, a \textit{quantum link formulation} can be adopted to map the gauge and field operators to spin-$S$ operators as
\begin{subequations}
    \begin{align}
        \hat{U}_{\ell,\ell+1}&\to\frac{\hat{s}^+_{\ell,\ell+1}}{\sqrt{S(S+1)}},\\
\hat{E}_{\ell,\ell+1}&\to g\hat{s}^z_{\ell,\ell+1},
\end{align}
\end{subequations}
which maps the commutation relations~\eqref{eq:comm} as
\begin{subequations}\label{eq:commp}
    \begin{align}
        \big[\hat{U}_{\ell,\ell+1},\hat{U}_{r,r+1}^\dagger\big]&\to\frac{2\delta_{\ell,r}\hat{s}^z_{\ell,\ell+1}}{S(S+1)},\\
\big[\hat{E}_{\ell,\ell+1},\hat{U}_{r,r+1}\big]&\to\frac{g\delta_{\ell,r}\hat{s}^+_{\ell,\ell+1}}{\sqrt{S(S+1)}}.
    \end{align}
\end{subequations}
Whereas the commutation relation~\eqref{eq:commB} is satisfied at any value of $S$, we see that commutation relation~\eqref{eq:commA} is satisfied in the limit of $S\to\infty$. In $1+1$D, the fermionic matter degrees of freedom can be mapped onto Pauli operators through a Jordan--Wigner transformation. This quantum link formulation of $1+1$D lattice QED gives rise to the spin-$S$ $\mathrm{U}(1)$ QLM~\eqref{eq:QLM}.

\end{shaded}

\twocolumngrid

To facilitate the following discussion, it is pertinent to put things on a formal footing and define relevant objects. Unless otherwise specified, we shall denote as $\hat{H}_0$ the LGT Hamiltonian, which we are interested in, and which we shall try to realize with a quantum simulator. The gauge symmetry of $\hat{H}_0$ is encoded in the commutation relations $\big[\hat{H}_0,\hat{G}_j\big]=0,\,\forall j$, where $\hat{G}_j$ is the \textit{generator} of the gauge symmetry. Since constituents in many common quantum simulators are naturally discrete, we focus here on LGTs, such that $j$ represents a given site on the lattice, though much progress has also been made in recent years in cold-atom simulations of field theories; see, e.g., \cite{Zache2020extracting,Pruefer2020experimental,Viermann2022quantum}. Intuitively, one can think of the operator $\hat{G}_j$ as a discretized version of Gauss's law from electrodynamics, see Fig.~\ref{fig:lgt_schematic}a,b. The dynamics of the gauge theory is thus determined by an extensive set of local conservation laws, one for each lattice site. For simplicity, and in line with current developments, we shall primarily focus on \textit{Abelian} LGTs, which host a single generator locally. In contrast, their \textit{non-Abelian} counterparts, which we will briefly discuss in the Perspective, host multiple generally noncommuting generators locally.

Gauge-invariant states are simultaneous eigenstates of the local generators: $\hat{G}_j\ket{\psi}=g_j\ket{\psi},\,\forall j$. A gauge superselection sector within the full many-body spectrum of the Hamiltonian $\hat{H}_0$ is defined by a unique set of these conserved eigenvalues $g_j$, so-called \textit{background charges}, over the entire system. On a quantum simulator, we are usually interested in restricting the physics to a single \textit{target} gauge superselection sector $g_j^\text{tar}$. In a quantum simulator, coherent errors can emerge that explicitly break the gauge symmetry, and they will be denoted by $\lambda\hat{H}_1$ with strength $\lambda$, where $\big[\hat{H}_1,\hat{G}_j\big]\neq0$. Processes due to $\lambda\hat{H}_1$ transition the system away from the target sector. These processes can be pure errors that deteriorate the dynamics, or they can represent interesting effects that generate new physics not present in the ideal LGT. 

Quantum link models (QLMs), widely used in quantum simulators, further discretize the continuous transporters of lattice quantum electrodynamics (QED) and QCD into quantum links, associated with discrete finite-dimensional local Hilbert spaces, while preserving the essential features of strong gauge dynamics. Our focus will be on the $1+1$D spin-$S$ $\mathrm{U}(1)$ QLM formulation~\cite{Chandrasekharan1997,Wiese_review} of $1+1$D lattice QED (see Box 1), given by the Hamiltonian~\cite{Kasper2017}
\begin{align}\nonumber
    \hat{H}_0=&-\frac{\kappa}{2a\sqrt{S(S+1)}}\sum_{\ell=1}^{L_\text{m}-1}\big(\hat{\sigma}_\ell^
    -\hat{s}^+_{\ell,\ell+1}\hat{\sigma}_{\ell+1}^-+\text{H.c.}\big)\\\nonumber
    &+\frac{\mu}{2}\sum_{\ell=1}^{L_\text{m}}\hat{\sigma}_{\ell}^z+\frac{g^2a}{2}\sum_{\ell=1}^{L_\text{m}-1}\big(\hat{s}^z_{\ell,\ell+1}\big)^2\\\label{eq:QLM}
    &+a\chi\sum_{\ell=1}^{L_\text{m}-1}(-1)^{\ell+1}\hat{s}^z_{\ell,\ell+1}.
\end{align}
In this formulation, the gauge and electric fields at the link between sites $\ell$ and $\ell+1$ are represented by the spin-$S$ operators $\hat{s}^+_{\ell,\ell+1}/\sqrt{S(S+1)}$  and $\hat{s}^z_{\ell,\ell+1}$, respectively. The matter field on site $\ell$ is represented by the Pauli operator $\hat{\sigma}^z_\ell$, where $\mu$ is the fermionic mass, $L_\text{m}$ is the number of sites, $\kappa$ is the tunneling strength, $g$ is the gauge coupling, $\chi=g^2(\theta-\pi)/(2\pi)$ quantifies the deviation of the topological $\theta$-angle from $\pi$, and $a$ is the lattice spacing, which we shall henceforth set to unity. Note that we have employed a particle--hole transformation to obtain the final form of Hamiltonian~\eqref{eq:QLM} \cite{Hauke2013}.

The $\mathrm{U}(1)$ gauge symmetry of Hamiltonian~\eqref{eq:QLM} is generated by the operator
\begin{align}\label{eq:G}
    \hat{G}_\ell=(-1)^\ell\bigg(\frac{\hat{\sigma}^z_\ell+1}{2}+\hat{s}^z_{\ell-1,\ell}+\hat{s}^z_{\ell,\ell+1}\bigg),
\end{align}
the eigenvalues $g_\ell$ of which take on the values $(-1)^\ell g_\ell\in\{-2S,-2S+1,\ldots,2S+1\}$. Typically, we want to work in the \textit{target} gauge superselection sector $g_\ell=0,\,\forall\ell$, also sometimes called the sector of Gauss's law.

In the limit of $S\to\infty$ at fixed lattice spacing $a$, Hamiltonian~\eqref{eq:QLM} retrieves lattice QED. Ideally, we want to aim to implement this Hamiltonian on a quantum simulator for as large $S$ as possible, but even small values of $S$ can give rise to very interesting physics relevant to QED \cite{Wiese_review}, such as the Coleman phase transition \cite{Coleman1976,Pichler2016} and confinement \cite{Surace2020}. Furthermore, the QLM is amenable for implementation in cold-atom setups, as we will detail in the following; see Fig.~\ref{fig:lgt_schematic}d.

\section{Quantum simulation with cold atoms}\label{sec:QSCA}
Cold neutral atoms offer a promising approach to create a scalable many-body system with controlled interactions~\cite{Lewenstein_book,Gross2017}. By cooling these atoms to ultra-low temperatures and trapping them optically, precise control over the system's geometry and dimensionality can be achieved. These optical traps create discretized lattice structures, similar to the discrete space in LGTs; see Fig.~\ref{fig:lgt_schematic}c,d. The interactions between the atoms can be adjusted to be stronger than the kinetic energy, allowing for exploration of the nonperturbative regime. Techniques such as optical lattices, Feschbach resonance, and long-range Rydberg interactions can facilitate this tuning. To represent fermion matter and gauge fields in LGTs, one can utilize fermionic or bosonic cold atoms, as well as different atom species. Ultracold atoms in optical lattices, described by the Hubbard model, provide a valuable platform for studying various quantum phases. Specifically, atomic Mott insulators serve as defect-free systems, which constitute an ideal starting point for high-fidelity initialization of the system for investigating strongly correlated phenomena in LGTs.

In this section, we will discuss recent technical advances in cold-atom experiments for simulating LGTs. Specifically, we will focus on techniques involving ultracold atoms in optical lattices, quantum gas microscopes, tweezer-trapped Rydberg-atom arrays, and emerging techniques utilizing alkaline-earth atoms. By exploring these advancements, we aim to uncover their potential applications in quantum simulations of LGTs and simplified QLMs, which nonetheless share many exciting properties with more complex gauge theories.

\begin{figure*}
	\centering
	\includegraphics[width=0.9\linewidth]{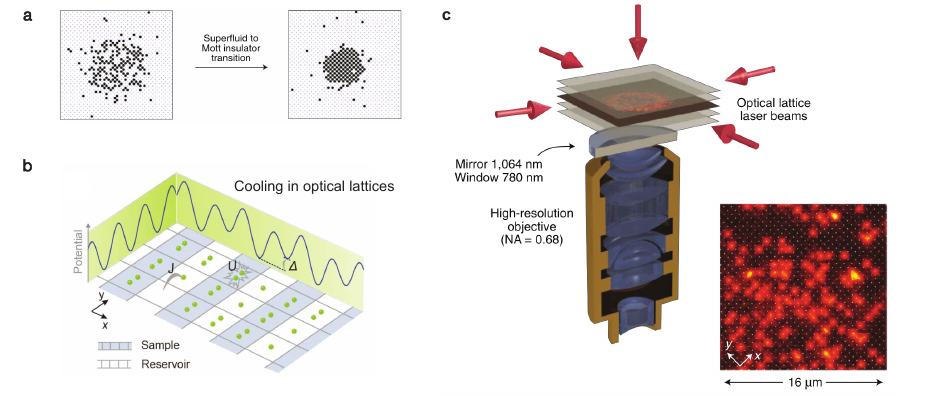}
	\caption{\textbf{Optical lattices and quantum gas microscopy.} 
    (a) Parity-projected site occupation in an optical lattice for a superfluid (left) with average occupation $n\simeq 1/2$ and for a Mott insulator (right) with average filling $n\lesssim 1$ (figure adapted from Ref.~\cite{Sherson2010}).
    (b) Schematic of the staggered-immersion cooling. A bichromatic optical superlattice is used to prepare low-entropy Mott insulators (samples) simultaneously with gapless superfluid reservoirs in a staggered configuration by introducing a potential energy offset $\Delta$ on every other site (figure adapted from Ref.~\cite{Yang2020Cooling}).
    (c) Illustration of a quantum gas microscope setup. Atoms are confined in a single plane of a vertical lattice. For fluorescence imaging atoms are pinned in a deep 3D lattice (red arrows). The fluorescence photons are collected with a high-NA imaging system. The signal from each atom (red) corresponds to the point-spread-function of the imaging system (figure adapted from Ref.~\cite{Sherson2010}).}
	\label{fig:lattice}
\end{figure*}

\subsection{Optical lattices}

When neutral atoms are cooled to temperatures in the nano-Kelvin range, their quantum properties become apparent, allowing for the creation of a scalable quantum system comprising up to $10^6$ identical particles~\cite{Anderson1995,Davis1995}.
By using spatially ordered optical potentials, the atoms can be distributed and rearranged into regular patterns.
These optical micro-traps offer precise control over the atomic states~\cite{Lewenstein_book}.
In particular, loading a quantum degenerate gas into such an optical lattice can trigger a quantum phase transition from a superfluid to a Mott insulator (MI)~\cite{Jaksch1998,Greiner2002,Jaksch2005}.
This transition is primarily driven by the competition between the kinetic energy that scales with the tunnel coupling between neighboring sites $J$ and the on-site Hubbard interaction energy $U$, which denotes the energy cost of having two particles on one lattice site (Fig.~\ref{fig:lattice}a).
In the strongly interacting limit where $J \ll U$, the phase transition leads to a uniformly filled MI~\cite{Bakr2010,Sherson2010}.
The MI at filling $n=1$, i.e., one atom per site, can be considered a quantum register, where various techniques for local state-dependent control of the atoms have been developed in order to realize well-defined initial states with high fidelity~\cite{Gross2017,Gross2021}.

The achievement of a defect-free MI can be considered as the initialization step in quantum simulation.
However, defects in the form of particle or hole excitations tend to increase with temperature~\cite{Trotzky2010,Sherson2010,Endres2011}.
While cooling of bulk gases can be achieved with high efficiency, the process of lattice loading and subsequent state manipulation typically introduces additional heating and the number of defects in the MI serves as a thermometer for the system.
Decreasing the initial-state temperature is one of the main challenges for high-fidelity quantum simulation experiments~\cite{McKay2011}.

Ultracold atoms confined in optical lattices are well isolated from the surrounding environment.
However, the cooling of such systems necessitates dissipative processes, which requires coupling with reservoirs to facilitate the cooling process. Developing these techniques for atoms in lattices is an outstanding challenge.
In an open system, it is crucial to identify a phase characterized by a small energy gap and a large density of states to absorb entropy from the gapped phase that one aims to cool.
One option following the theoretical ideas presented in Ref.~\cite{Ho2009} constitutes extracting entropy by utilizing a surrounding superfluid or metallic reservoir. Similar ideas could be realized in a bilayer system~\cite{Kantian2018}.
The former has been successfully implemented in Ref.~\cite{Mazurenko2017} to reveal anti-ferromagnetic correlations in the low-temperature phase of the Fermi-Hubbard model at half filling. One of the main limitations of the idea is the scaling of the cooling performance with system size, since the contact between the gapped and gapless phase exists only at the circumference of the central region which hosts the gapless low-entropy phase.
Using superlattices, however, superfluid reservoirs can be immersed as stripes within $1$D bosonic MI chains (Fig.~\ref{fig:lattice}b) as demonstrated in Ref.~\cite{Yang2020Cooling}.
By making use of the large density of states in the superfluid phase, the superfluid reservoirs can efficiently extract entropy from the system.
Subsequently, a state-engineering technique was employed to achieved a unity-filled MI with only $0.8(1)\%$ of defects.
This experimental implementation provides an excellent platform for the quantum simulation of various models, for instance the realization of a bosonic antiferromagnet state~\cite{Sun2021}.

\subsection{Quantum gas microscopy}

The quantum states of cold atoms offer a rich set of degrees of freedom that can be utilized to encode the fermionic matter and dynamical gauge field in LGTs. Within the framework of the QLM formalism, spins can be effectively represented by the internal energy levels of atoms~\cite{Buechler2005,Dai2017}, the occupation numbers on lattice sites~\cite{Cirac2010,Schweizer2019,Yang2020}, and different atomic species ~\cite{Banerjee2012,Zohar2013,Zache2018,Mil2020}. In addition, quantum gas microscopes (Fig.~\ref{fig:lattice}c) facilitate the detection and manipulation of individual atoms 
with extreme resolution down to the level of a single lattice site~\cite{Gross2021}. 
Following early pioneering works on single-atom sensitive imaging~\cite{Schlosser2001,Miroshnychenko2003,Nelson2007}, 
the advent of quantum gas microscopes in 2009 
enabled detection fidelities near unity of dense low-entropy 
strongly correlated many-body systems~\cite{Bakr2009,Sherson2010}.

The detection method is based on fluorescence imaging, where individual atoms
scatter several thousand photons, which are then collected with a high-resolution
imaging system (Fig.~\ref{fig:lattice}c). Typical length scales are given by the lattice constant and 
are on the order of or less than the optical wavelength used to generate the potential.
Reaching an optical resolution on the order of $\sim 0.5\,\mu$m requires careful 
engineering of the experimental apparatus, e.g., by designing high-NA imaging
objectives~\cite{Robens2017} or via two-dimensional tunable accordion lattices~\cite{Su2023}. Combining microscope setups with additional experimental techniques specific to the respective atomic species requires the construction of elaborate experimental setups~\cite{Sohmen2023}.
At the same time, a large signal-to-noise ratio
is essential. This can even aid high-fidelity reconstruction of the density distribution
in regimes, where naively the optical resolution would only be able to distinguish particles
at a distance of more than two lattice constants~\cite{Impertro2023}.
The challenges associated with a large number of scattered photons are atom loss and temperature-assisted
tunneling processes, i.e., while the atoms scatter photons, their temperature in the
lattice well increases and they can move to a different lattice site further away,
where they are recaptured, or they can be completely lost from the trap. 
Both processes result in false detection events and need to be minimized 
by applying efficient cooling during imaging and by pinning the atoms in very
deep optical lattices. While the first quantum gas microscopes have been developed
for bosonic Alkali atoms~\cite{Bakr2009,Sherson2010} due to the existence of less complex 
imaging and cooling techniques, there is now a large number of
microscopes for fermionic Alkali atoms~\cite{Cheuk2015,Haller2015,Parsons2015,Edge2015}, and recent developments have further extended
their application to Alkaline-earth(-like) atoms~\cite{Miranda2015,Yamamoto2016} 
as well as magnetic atoms~\cite{Su2023} and even molecules~\cite{Rosenberg2022,Christakis2023}.
Quantum gas microscopes offer unique opportunities
to study the properties of quantum many-body systems
via direct detection of entanglement entropy~\cite{Islam2015,Kaufman2016,Lukin2019,Rispoli2019}, 
string order~\cite{Endres2011,Chiu2019}, 
hidden order~\cite{Hilker2017}, 
(multi-point) correlation functions~\cite{Cheneau2012,Zheng2022}, and full counting statistics~\cite{Wei2022,Wienand2023}.

High-resolution imaging also offers intriguing possibilities for local control and manipulation 
at the level of a single atom~\cite{Weitenberg:2011}, which is enabled by precise shaping of optical potentials at length scales on the order of $0.5\,\mu$m~\cite{Zupancic2016}. Starting from an initial MI at unit filling,
the internal state of the atom can be rotated by introducing light that couples to a different internal state. Addressability is provided by applying local differential light shifts in combination with a global coupling laser or microwave field~\cite{Rubio-abadal2019}. Moreover, by employing resonant light, atoms in internal states that are sensitive to it can be heated out of the trap. This enables the preparation of essentially arbitrary initial product states. This method is sometimes also referred to as ``cookie cutting'' in contrast to local rearrangement techniques facilitated by
tightly focused movable tweezer beams~\cite{Young2022}, as we will discuss in more detail below.
Local addressability has been particularly useful for studying few-particle systems, such as magnon excitations in spin chains~\cite{Fukuhara2013}, spin-charge separation in $1$D Fermi-Hubbard chains~\cite{Vijayan2020}, few-particle dynamics~\cite{Preiss2015,Tai2017}, and $1$D anyons~\cite{Kwan2023}, to name only a few. Moreover, state-dependent superlattices provide additional tools for preparing high-fidelity initial states~\cite{Yang2020Cooling}. Initial states that are prepared with this technique share the same translation symmetry as the lattice; hence, this scheme is less programmable, but it can give rise to higher fidelities due to the relative stability between the lattice and the local state-dependent light shifts.
In the context of the quantum simulation of LGTs, local addressing
for precise initial-state preparation is particularly important, since studying
a certain physical subspace of the many-body spectrum, such as given by the superselection sectors ${g_j}$,  relies on the preparation of specific configurations of particles in the initial state.

\begin{figure*}
	\centering	\includegraphics[width=0.9\linewidth]{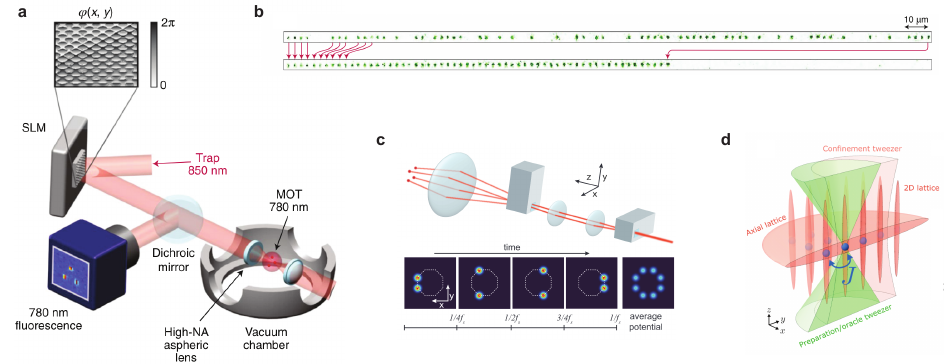}
	\caption{\textbf{Optical tweezer arrays.} 
    (a) Illustration of a tweezer array platform, where an SLM imprints is used to generated tweezer arrays via application of a phase mask on the SLM and a high-NA imaging system (figure adapted from Ref.~\cite{Nogrette2014}). 
    (b) $1$D array of optical tweezers generated with an acousto-optical deflector (AOD) and multiple radio-frequency tones. After detecting the position of the atoms the traps are rearranged in order to form a defect-free array (figure adapted from Ref.~\cite{Endres2016}).
    (c) Illustration of a $2$D AOD setup in $4f$-configuration for tweezer generation. By stroboscopically switching between different configurations, the atoms effectively see an average potential in the form of a ring that consists of eight individual traps, which are close enough such that tunneling is significant for light Li atoms (figure adapted from Ref.~\cite{Yan2022}).
    (d) Schematic of a hybrid tweezer-lattice setup. Atoms are first loaded and cooled in optical tweezers (preparation/oracle tweezer, green), and then implanted into a 3D optical lattice (red), which consists of a vertical $1$D lattice to confine the atoms in a single plane and a $2$D square horizontal lattice. The large-waist tweezer (confinement tweezer, pink) can be used to introduce an additional in-plane harmonic confinement (figure adapted from Ref.~\cite{Young2022}).}
	\label{fig:tweezer}
\end{figure*}

\subsection{Alkaline-earth atoms}
The rich internal level structure of Alkaline-earth(-like) atoms, such as Sr or Yb, significantly extends the available toolbox for state manipulation and detection~\cite{Takahashi2022}. These atoms have two valence electrons. Hence, the spectrum consists of a singlet and triplet manifold with the absolute ground state being the singlet state $^1$S$_0$. There are several ultra-narrow transitions that connect the ground state with triplet states~\cite{Hoyt2005,Yamaguchi2010,Ishiyama2023,Kawasaki2023}, which are of importance for applications in quantum simulation, computation, and the study of fundamental physics. The most prominent transition is $^1$S$_0$-$^3$P$_0$~\cite{Yamaguchi2010}, which is used in optical lattice clocks~\cite{Ludlow2015}. The lifetime of the excited states is many seconds, such that atoms prepared in the excited clock state are stable for the duration of typical experiments and can be viewed as a second species. This offers the possibility for implementing highly tunable state-dependent optical potentials with low scattering rates~\cite{Riegger2018,Heinz2020,Oppong2022,Hohn2023}. In combination with local control, new methods for engineering Gauss's law come within reach. This has recently been worked out in a theoretical proposal for studying $\mathrm{U}(1)$ QLMs in one and two dimensions~\cite{surace2023abinitio}. 

Alkaline-earth(-like) atoms further exhibit exotic interactions, which provide a fruitful playground for the design of more complex LGTs. The fermionic isotopes of Sr and Yb have a nonzero nuclear spin $I$, and the interactions between them are SU($N$) symmetric, i.e., independent of the internal state, where $N=2I+1$~\cite{Scazza2014,Zhang2014}. While the conventional SU(2) Fermi-Hubbard model (FHM) has attracted a lot of interest, as it has been suggested as a minimal model for high-$T_{\text{c}}$ superconductivity, the SU($N$) FHM is expected to host even more exotic quantum phases, such as chiral spin liquids~\cite{Cazalilla2014}. Moreover, the long lifetime of the excited clock state $^3$P$_0$ facilitates the realization of two-orbital Hubbard models, where spin-exchange plays an important role. In particular, for fermionic $^{173}$Yb the ferromagnetic exchange interaction is extremely large in contrast to typical spin-exchange interaction strengths that emerge within second-order perturbation theory~\cite{Trotzky2008}. While the exchange interaction for fermionic $^{87}$Sr is also ferromagnetic, it is of antiferromagnetic nature for $^{171}$Yb providing a route towards studying the Kondo lattice model~\cite{Doniach1977} (mappable onto an effective LGT problem) and Ruderman-Kittel-Kasuya-Yoshida interactions~\cite{Ruderman1954}. Harnessing the exotic interactions of Alkaline-earth(-like) atoms appears to be a promising direction towards quantum simulation of non-Abelian LGTs \cite{Halimeh2023Spin}.

The rich level structure of Alkaline-earth(-like) atoms further offers the possibility of implementing partial measurements, which could be interesting for the preparation of exotic initial states potentially with large entanglement~\cite{Mamaev2019}. Being able to prepare different types of physical initial states is an important ingredient for quantum simulation of LGTs. In addition, implementing fast imaging schemes enables feedback. This was recently demonstrated in tweezer arrays, where mid-circuit measurements have been implemented~\cite{Ma2023,Lis2023,Norcia2023}. Extending these ideas to optical lattices is possible, although technically more demanding due to the lower depth and risk of thermally activated tunneling during the imaging process.

\subsection{Rydberg-atom arrays}

Atoms trapped in optical tweezers have recently emerged as an alternative platform for quantum technologies~\cite{Saffman2010,Browaeys2020,Adams2019,Wu2021}. Here, instead of the regular interference pattern characterizing standing-wave optical lattices, individual atoms are held in tightly focused optical tweezer beams, which are typically separated by a few $\mu$m. The tweezer array is usually either generated with a spatial light modulator (SLM), as illustrated in Fig.~\ref{fig:tweezer}a, or acousto-optic modulators (AODs) with multiple radio-frequency tones (Fig.~\ref{fig:tweezer}b,c). An advantage of tweezer arrays is that this setup eliminates the need to cool the atoms to the quantum degenerate regime, which significantly reduces experimental complexity and cycle times. Moreover, since the tweezer array can be produced with SLMs, the topology of the array can be freely chosen in contrast to optical lattices. 
In these setups, atoms are directly loaded into the array after a laser cooling stage. Due to light-assisted collisions, each trap is at most occupied by one atom with a probability of $p\simeq 1/2$~\cite{Schlosser2001}. This results in an initial state where the atoms are stochastically loaded into the traps and hence appear in random locations. In order to remove entropy and obtain a uniform array with precisely one atom per trap, the atoms need to be rearranged~\cite{Kim2016,Endres2016,Barredo2016}. To this end, the atoms are imaged after the initial loading step using nondestructive fluorescence imaging. Based on the measured occupations, atoms will be moved with dynamically configurable optical tweezer beams to their final locations in order to produce defect-free arrays (Fig.~\ref{fig:tweezer}b). 

In contrast to optical lattice experiments, there is no tunneling between neighboring traps. Instead, interactions are induced by exciting the atoms to highly excited Rydberg states~\cite{Saffman2010}. The interaction strength can be adjusted either by controlling the interatomic distance or by utilizing different Rydberg states. The large dipole-dipole interaction between Rydberg atoms can be employed to implement different spin models. Here, the ground $\ket{g}$ and Rydberg  $\ket{r}$ states are mapped onto a spin-$1/2$ system, and the van der Waals interaction between Rydberg atoms gives rise to an Ising-type spin Hamiltonian whose interactions scale as $1/R^6$ as a function of the separation $R$ between atoms. In addition, the atoms are driven by a laser beam that couples $\ket{g}$ and $\ket{r}$. Here, the coherent coupling maps to a transverse field while the detuning from resonance maps to a longitudinal field~\cite{Browaeys2020}. All parameters are independently tunable either by changing the parameters of the coupling laser or by adjusting the distance between neighboring atoms. Moreover, using local addressing, the Hamiltonian parameters can be adjusted on the level of individual atoms~\cite{Labuhn2014}. Coherent many-body dynamics and phase transitions have been studied for such an Ising-type Hamiltonian with up to $51$ atoms in Ref.~\cite{Bernien2017}.

Using a system of two Rydberg states $\ket{r_1}$ and $\ket{r_2}$, the resonant dipole-dipole interaction maps onto an XY-spin model, where transverse and longitudinal fields are realized with a microwave field. The two-dimensional dipolar XY-model, for instance, served as a basis for the observation of scalable spin squeezing in Rydberg arrays~\cite{Bornet2023}, which has also been achieved with Ising-type interactions on an optical clock transition~\cite{Eckner2023}. Moreover, making use of Rydberg blockade between nearest neighbors, dimer models can be efficiently implemented, which for a filling of $1/4$ on a Kagom\'e lattice facilitated the preparation of nontrivial states with spin liquid character~\cite{Semeghini2021}. Rydberg atom arrays are further ideally suited for studying hybrid digital-analog or purely digital quantum information protocols. This has been demonstrated in Ref.~\cite{Bluvstein2022} where coherent transport of entangled atoms was used to realize highly entangled states including the toric code state on a torus with sixteen data and eight ancillary qubits. 

\subsection{Hybrid-tweezer lattices}
One of the main challenges for the reliable quantum simulation of LGTs is the careful engineering of local symmetries, so that Gauss's law is respected even for long evolution times. This does not occur naturally in Hubbard-type optical lattices, where symmetries are typically global. A promising route to overcome these limitations constitutes combining optical lattices with local control provided by optical tweezer arrays. While it is technically challenging to combine both setups in a way that provides good relative stability in terms of the position of the tweezer beams relative to the lattice sites, there have been significant experimental advances recently. One pathway towards enhancing the programmability of atom arrays with significant tunneling between neighboring sites is to bring the tweezer traps close enough for tunneling to occur. The main challenge here are position and intensity fluctuations, which will result in decoherence and dephasing of tunneling between neighboring traps. Following early pioneering work with Rb atoms, these techniques have been extended to two dimensions for experimental studies of programmable Fermi-Hubbard systems of a few particles~\cite{Spar2022,Yan2022}, as illustrated in Fig.~\ref{fig:tweezer}c. These implementations are based on light Li atoms, where larger separations between potential wells can still result in sizable tunnel couplings. A second pathway consists in the use of local potentials to locally manipulate the potential energy of selected sites in order to detune or suppress tunneling between neighboring sites (Fig.~\ref{fig:tweezer}d) as recently demonstrated with bosonic Sr atoms~\cite{Young2022,Tao2023} and in a Rb quantum gas microscope~\cite{Wei2023}. Combining optical lattices with optical tweezers further offers intriguing possibilities for increasing the repetition rate of quantum simulation experiments via direct laser cooling of atoms in tightly focused tweezer traps, which can then be used to implant atoms into an optical lattice at the desired positions~\cite{Young2022}. In the context of LGTs, where dynamics is one of the most immediate applications of quantum simulation, this approach provides an exciting route for the realization of specific physical initial states.

\section{First steps and building blocks}\label{sec:BuildingBlocks}
\begin{figure*}
	\centering	\includegraphics[width=0.99\linewidth]{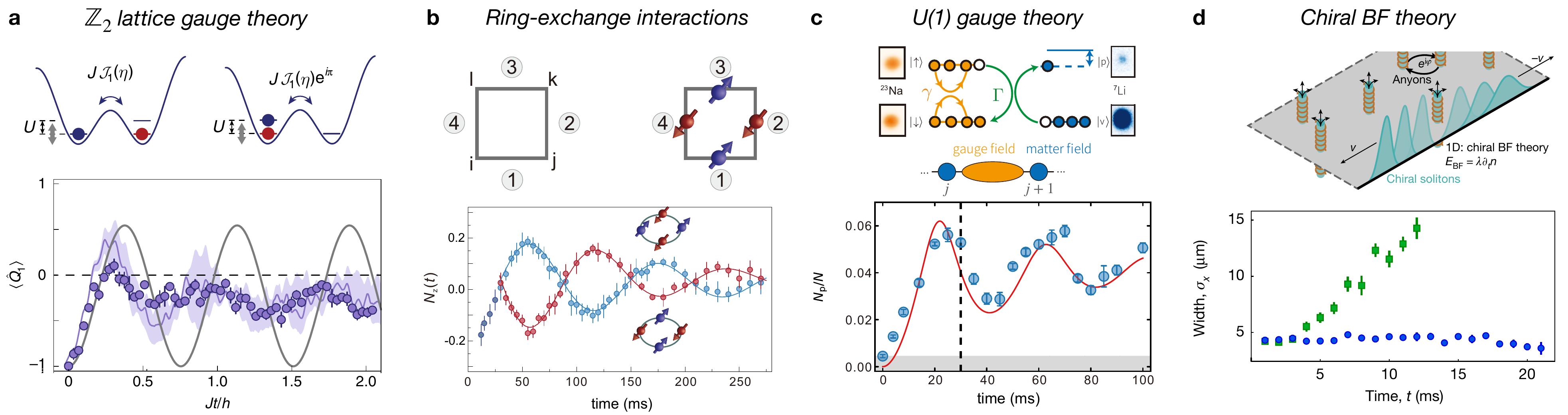}
	\caption{\textbf{First steps and building blocks.} Cold atom experiments have realized various building blocks of gauge theories already: 
 (a) A $\mathbb{Z}_2$-minimal coupling of matter with a $\mathbb{Z}_2$ gauge field was realized through Floquet engineering~\cite{Schweizer2019}. Two atomic hyperfine states (red, blue) on the sites of a double-well realize matter and gauge fields respectively. The experiment \cite{Schweizer2019} observed a dynamical redistribution of $\mathbb{Z}_2$ charge $\langle \hat{Q}_1 \rangle = \langle e^{i \pi \hat{n}^a_1} \rangle$ (bottom panel).  (Figure adapted from~\cite{Schweizer2019}.)
 (b) Ring exchange interaction between four spins was achieved by coupling cold atoms in an optical lattice through fourth-order perturbation processes. The interaction induces oscillations between the states $\ket{\uparrow\downarrow\uparrow\downarrow}$ and $\ket{\downarrow\uparrow\downarrow\uparrow}$, as observed in the experiment (bottom panel)~\cite{Dai2017}. The reversal of the oscillation at the quarter period was measured, providing evidence for the anyonic statistics exhibited by the system. (Figure adapted from~\cite{Dai2017}.)
 (c) A $\mathrm{U}(1)$-minimal coupling of matter with a $\mathrm{U}(1)$ gauge field was realized using a two-species mixture of sodium (blue) and lithium (orange) atoms~\cite{Mil2020}. Matter fields occupy sites $...,n, n+1,...$ and gauge fields the links in-between. Heteronuclear spin-changing collisions produce the dynamics corresponding to a gauge-invariant coupling between these two fields, retaining $\mathrm{U}(1)$ gauge symmetry. The dynamics of particle production is shown (bottom panel), where the evolution of the measured density distribution $N_\text{p}/N$ in state $\ket{\text{p}}$ is plotted.  (Figure adapted from~\cite{Mil2020}.)
 (d) A chiral BF theory, a one-dimensional reduction of Chern-Simons theory, was realized through Raman coupling after integrating out the gauge field~\cite{Froelian2022}. The experiment observed a chiral soliton, which can only propagate robustly along one direction (bottom panel).  (Figure adapted from~\cite{Froelian2022}.)}
	\label{Fig3}
\end{figure*}

Cold-atom experiments have already made significant steps along the way towards simulating full-blown gauge theories with couplings to dynamical matter. Several implementations so far have focused on realizing individual building blocks, demonstrating local gauge-invariant couplings on the smallest possible scales, see Fig.~\ref{Fig3}. Another route starts by formally integrating out the gauge fields, which leads to nontrivial couplings between the remaining matter fields which can also be realized experimentally. In the following, we review the key experiments so far involving cold atoms, all of which could potentially be scaled up to realize extended gauge theories.

\subsection{$\mathbb{Z}_2$ gauge theory by Floquet engineering}
The simplest representation of a gauge field corresponds to a gauge degree of freedom on the links of a lattice with a two-dimensional Hilbert space, which we denote by $\hat{\tau}^z_{\ij}$ in the following; it takes the role of a parallel transporter $\hat{U}_\ij$ on link $\ij$. The conjugate electric-field variable can be described by the Pauli matrix $\hat{\tau}^x_\ij$, with its positive ($\ket{+x}$) and negative ($\ket{-x}$) eigenvectors corresponding to the allowed electric-field configurations. Different proposed implementations work in different bases, typically either the $\hat{\tau}^x$ or $\hat{\tau}^z$ eigenbasis is used.

Out of such spin-$1/2$ gauge degrees of freedom, different local symmetries $\hat{G}_{\vec{j}}$ can be constructed such that $[\hat{H}_0 , \hat{G}_{\vec{j}}] =0$. These define different local gauge constraints, i.e., different Gauss's laws: If the adjacent link variables $\hat{\tau}^x_\ij$ are combined in a sum, the spin-1/2 $\mathrm{U}(1)$ QLM is obtained. If, on the other hand, products of the electric fields are taken, $ \hat{G}_{\vec{j}} \propto \prod_{\vec{i}:\ij} \hat{\tau}^x_\ij$, the \emph{$\mathbb{Z}_2$ gauge group} is obtained; here the product includes all bonds $\ij$ including the reference site $\vec{j}$.

In addition to the link variables, matter fields $\a_{\vec{j}}$ can be defined on the sites $\vec{j}$ of the lattice. The simplest case assumes hard-core bosons $\a_{\vec{j}}$ whose parity corresponds to their $\mathbb{Z}_2$ charge. I.e., the local $\mathbb{Z}_2$ gauge symmetry becomes $[\hat{H}_0, \hat{G}_{\vec{j}}]=0$, with
\begin{equation}
    \hat{G}_{\vec{j}} = (-1)^{\hat{n}_{\vec{j}}}  \prod_{\vec{i}:\ij} \hat{\tau}^x_\ij, \quad \hat{n}_{\vec{j}} = \ad_{\vec{j}} \a_{\vec{j}}.
\end{equation}
The corresponding Gauss's law constraint reads
\begin{equation}
    \hat{G}_{\vec{j}} \ket{\Psi} = g_{\vec{j}} \ket{\Psi}, \qquad g_{\vec{j}} \in \{ -1,+1 \},
\end{equation}
where $g_{\vec{j}}$ describes a given configuration of background charges. For a lattice without open links, these have to satisfy $\prod_{\vec{j}} g_{\vec{j}}= 1$, since $\prod_{\vec{j}} \hat{G}_{\vec{j}}=\mathds{1}$.

The simplest $\mathbb{Z}_2$ gauge-invariant term, besides $\hat{\tau}^x_{\ij}$ and $\hat{n}_{\vec{j}}$, that may appear in the Hamiltonian corresponds to the \emph{$\mathbb{Z}_2$ minimal coupling}:
\begin{equation}
    \hat{H}_{\rm min, \mathbb{Z}_2} = - J_a ~ \ad_{\vec{i}} \hat{\tau}^z_{\ij} \a_{\vec{j}} + \hc
    \label{eqHminZ2}
\end{equation}
It describes the elementary tunneling process of a hard-core boson between neighboring sites $\vec{i}$ and $\vec{j}$. Formulated in the eigenbasis of $\hat{\tau}^z_{\ij}$, the sign of the tunneling matrix element is determined by the gauge field; formulated in the eigenbasis of $\hat{\tau}^x_\ij$, the string configuration is flipped from $\ket{\pm x}$ to $\ket{\mp x}$ upon tunneling.

The minimal instance of the $\mathbb{Z}_2$ gauge theory described above, with one matter particle (denoted by $a$) hopping between two sites with a gauge link in between, was realized by Schweizer \textit{et al.}~\cite{Schweizer2019} following a theoretical proposal by Barbiero \textit{et al.}~\cite{Barbiero2019}. The experiment utilized a well-known trick for engineering synthetic gauge fields for cold atoms~\cite{Jaksch2003,Aidelsburger2011}, where tunneling is first suppressed by a strong potential gradient and subsequently restored by a resonant modulation whose phase $\phi$ determines the resulting hopping (or Peierls) phase. 

As shown in the top panel of Fig.~\ref{Fig3}a, Schweizer \textit{et al.}~\cite{Schweizer2019} replaced the strong potential gradient by a large on-site Hubbard interaction $U \gg J$ with a second atom (denoted by $f$) in a different hyperfine state; $J$ denotes the bare tunneling amplitude. As a consequence, the acquired phase of the restored tunneling matrix element depends on the configuration of the $f$ atom~\cite{Bermudez2015}. This procedure allows to make the synthetic gauge field experienced by the $a$-particle dependent on the density of the $f$-particle, and has been utilized more broadly to realize density-dependent gauge fields~\cite{Clark2018PRL}.

Schweizer \textit{et al.}~\cite{Schweizer2019} went a step further by identifying the spatial degrees of freedom of the $f$-particle with the eigenstates of $\hat{\tau}^z_\ij$. Noting that a relative $\pi$-phase shift occurs in the sign of the restored tunneling matrix element of the matter particle $a$ when the latter hops onto / off of a site occupied by the $f$-particle, the desired minimal coupling Eq.~\eqref{eqHminZ2} is realized. The tunneling amplitude is renormalized by the first Bessel function, $J_a = J \mathcal{J}_1(\eta)$, where $\eta$ is the amplitude of the potential modulation in units of the resonant shaking frequency $\hbar \omega = U$.

In order to introduce nontrivial dynamics of the gauge field itself, Schweizer \textit{et al.}~\cite{Schweizer2019} realized the electric-field term in their Hamiltonian,
\begin{equation}
    \hat{H}_{\rm el, \mathbb{Z}_2} = - J_f \hat{\tau}^x_\ij,
    \label{eqHelZ2}
\end{equation}
by further suppressing tunneling of the $f$-particle through a magnetic gradient $\Delta$ seen only by the $f$-particle. For a proper choice of the shaking amplitude $\eta=1.84$, for which $\mathcal{J}_0(\eta) = \mathcal{J}_2(\eta) = J_f/J$, they demonstrated that the restored tunneling of the $f$-particle --- corresponding to a spin-flip in the eigenbasis of $\hat{\tau}^z_\ij$ --- becomes independent of the $a$-configuration, as required by Eq.~\eqref{eqHelZ2}.

As a demonstration of their quantum simulator, Schweizer \textit{et al.}~\cite{Schweizer2019} initialized an $a$-particle on the left of the double-well and an $f$-particle in an eigenstate of $\hat{\tau}^x_{\ij}$. The subsequent dynamics of the charge operator on this site, $\langle \hat{Q}_1 \rangle = \langle e^{i \pi \hat{n}^a_1} \rangle$, is shown in the bottom panel of Fig.~\ref{Fig3}a. It shows oscillations, which are damped due to gauge-non invariant processes $\hat{H}_1$ originating from averaging over multiple double wells in an inhomogeneous trapping potential. 

Overall, the experiment demonstrates the principle feasibility of simulating $\mathbb{Z}_2$ LGTs coupled to dynamical matter. Theoretical analysis by Schweizer \textit{et al.}~\cite{Schweizer2019} suggests that gauge invariance can be ensured in the building block by fine-tuning the experimental parameters. Accompanying theoretical work~\cite{Barbiero2019} demonstrated the scalability of the approach to extended one-dimensional systems if one utilizes two-frequency driving to engineer density-dependent Peierls phases. The latter were demonstrated in a parallel experiment by G\"org \textit{et al.}~\cite{Goerg2019}, completing the demonstration of all required ingredients to realize larger-scale $\mathbb{Z}_2$ LGTs in contemporary cold-atom setups. Further extensions in Rydberg-atom tweezer arrays were recently proposed theoretically by Homeier \textit{et al.}~\cite{Homeier2023}.

\subsection{Ring exchange from high-order perturbation theory}
The magnetic field in gauge-theory quantum simulators only appears beyond one spatial dimension and represents a particular challenge due to its multiparticle/higher-order nature. In higher-dimensional lattice QED, such as $2+1$D, the magnetic-field operator can be represented by the \emph{plaquette} term~\cite{Wilson1974,Kogut1975,Wiese2013}.
As illustrated in Fig.~\ref{Fig3}b, the plaquette is a small square created by connecting four neighboring lattice sites.
The links of this plaquette are identified by indices $n$ ranging from $1$ to $4$, and each link is associated with a dynamical operator denoted as $\hat{U}_n$ or $\hat{U}_{ij}$ from site $i$ to $j$.
This gauge-invariant term is a product of four link variables $\hat{U}_n$, which can be written as
\begin{equation}
\hat{U}_{\Box} = \hat{U}_{ij} \hat{U}_{jk} \hat{U}_{lk}^{\dagger} \hat{U}_{il}^{\dagger},
\end{equation}
and it flips the electric fields around the edge of a loop. 
This plaquette term represents the simplest nontrivial form of a Wilson loop, which involves a closed loop path integral of the gauge fields.
In the QLM formalism, the term in the Hamiltonian associated with the plaquette operator manifests itself as the ring-exchange interactions among four spins.
These interactions can be described as follows for a square plaquette, 
\begin{equation}
\hat{H}_{_{\text{Ring}}} = -J_{_\square} (\hat{S}_1^{+} \hat{S}_2^{-} \hat{S}_3^{+} \hat{S}_4^{-} + \text{H.c.}),
\end{equation}
where $\hat{S}^{\pm}_n$ represent the spin raising and lowering operators, respectively, on link $n$, and $J_{_\square}$ denotes the coupling strength of the four-body ring-exchange interaction; see Fig.~\ref{Fig3}b, which illustrates the action of the ring-exchange term.

The four-body ring-exchange interaction can be derived through a higher-order expansion of the Hubbard Hamiltonian~\cite{MacDonald1988}.
Therefore, ultracold atoms confined in optical lattices which are well-described by the Hubbard model, can serve as a suitable quantum simulator for investigating dynamics of gauge theories mediated by the ring-exchange interaction.
In Ref.~\cite{Buechler2005}, an initial proposal suggested coupling the spins of the four links using a `molecular' state.
Additionally, Ref.~\cite{Paredes2008} proposed the implementation of a ring-exchange interaction within a square optical superlattice.
Notably, they suggested a scheme for suppressing lower-order processes which would otherwise govern the spin dynamics of the plaquette system.

Experimental observation of the four-body ring-exchange interaction was first reported in Ref.~\cite{Dai2017}.
The technical developments on full control of the atomic states in square superlattices enable practical realization~\cite{Trotzky2008,Yang2017}.
Starting with a unity-filled MI state, the spin configuration was initialized to $\Ket{\uparrow \downarrow \uparrow \downarrow}$.
The high barriers between neighboring plaquettes isolate the system into an array of four-site building blocks.
In each plaquette, the bare tunneling dynamics is blocked by the on-site interaction $U$ as the system was turned into the strongly interacting regime with $J \ll U$.
Furthermore, the superexchange process that involves second-order interactions was suppressed by an effective magnetic gradient field.
Since the ring-exchange coupling strength $J_{_\square}\approx 40 J^4/U^3$ is very sensitive to the Hubbard parameters, the experiment was carried out in a nearly homogeneous regime.
Finally, as shown in Fig.~\ref{Fig3}b, with excellent controllability of quantum coherence, the ring-exchange oscillations between the state $\Ket{\uparrow \downarrow \uparrow \downarrow}$ and $\Ket{\downarrow \uparrow \downarrow \uparrow}$ were clearly measured, ranging from $3$ Hz up to $36$ Hz~\cite{Dai2017}.

The ability to generate such ring-exchange interactions opens a range of interesting perspectives. For example, it is a key ingredient of Kitaev's toric code model~\cite{Kitaev2003} in the square lattice.
While Kitaev's toric code is effectively a $\mathbb{Z}_2$ LGT without dynamical matter, in Ref.~\cite{Dai2017} the ring-exchange interaction was introduced and interpreted within the language of the toric code Hamiltonian. Namely, in the subspace realized experimentally ($\ket{\uparrow\downarrow\uparrow\downarrow}$ and $\ket{\downarrow\uparrow\downarrow\uparrow}$), the Hamiltonian takes the form $\hat{H}_\text{Ring}=-J_{_\square}\hat{\sigma}^x_1\hat{\sigma}^x_2\hat{\sigma}^x_3\hat{\sigma}^x_4$.
As shown in Ref.~\cite{Kitaev2003}, the anyonic statistics measured in this building block imply that topological matter could be engineered and quantum-simulated in this many-body system~\cite{Goldman2016}.
It is known that condensed matter systems with significant ring exchange interactions exhibit exotic phases such as quantum spin liquids~\cite{Hermele2004}, or high-$T_\text{c}$ superconductivity~\cite{Moessner2001,Balents2002}.
However, due to the relatively small interaction strength in the ultracold-atom simulators, observing novel quantum phases driven by the plaquette ring-exchange interaction remains a challenge.
To address this issue, enhancing the ring-exchange interaction becomes crucial.
By increasing the on-site interaction strength $U$ while maintaining the ratio $J/U$, the magnitude of $J_{_\square}$, which scales as $J^4/U^3$, will be amplified.

\subsection{$\mathrm{U}(1)$ gauge theory from angular momentum conservation}\label{sec:U1BuildingBlock}

In Ref.~\cite{Zohar2013PRA}, it was realized that by suitably separating degrees of freedom, the global symmetry of angular momentum conservation can be promoted to a local symmetry.  
This approach has been further extended, refined, and simplified in a series of theory works~\cite{Stannigel2014,Kasper2016,Kasper2017,Zache2018}, until it became possible to implement a first building block of a $\mathrm{U}(1)$ QLM in the large-$S$ limit in a two species cold-atom experiment.

The employed model and setup were as follows. 
As sketched in Fig.~\ref{Fig3}c, at each site of the optical lattice, two degrees of freedom reside, labeled by $\text{v}$ and $\text{p}$, and governed by annihilation (creation) operators $\hat{b}_{j,\text{v}}$ and $\hat{b}_{j+1,\text{p}}$  ($\hat{b}_{j,\text{v}}$ and $\hat{b}_{j+1,\text{p}}$), respectively. These will represent the matter of the theory. Within the Schwinger model, these degrees of freedom are fermionic, but in the experiments of Mil \emph{et al.} \cite{Mil2020} they were chosen as bosonic, getting close to the Abelian Higgs model \cite{Bazavov2015}. As the labeling indicates, two partners from neighboring optical-lattice sites are considered as sitting at the same matter site of the target LGT. In this way, this model realizes the upper and lower components of a Dirac spinor via so-called Wilson fermions \cite{Zache2018} rather than the more commonly used staggered fermions \cite{Kogut1975}. 
Within a matter site (across two sites of the optical lattice), the two components can be coupled by a laser-assisted tunneling of strength $\Omega$, described by the Hamiltonian $\hat{H}_\mathrm{l.a.t.} = \sum_j \hbar\Omega \hat{b}_{j,\mathrm{p}}^\dagger\hat{b}_{j,\mathrm{v}}+\mathrm{H.c.}$

An additional deeper optical lattice strongly traps a second bosonic atomic species described by two internal states $\sigma=0,1$ and with associated annihilation (creation) operators $\hat{a}_{\sigma,j}$ ($\hat{a}_{\sigma,j}^\dagger$). 
Locally, we can identify these bosonic fields via the Schwinger representation with spin operators of length,  
\begin{subequations}
	\begin{align}
	\hat{L}^{+}_{j,j+1} &= \hat{a}_{0,j}^\dagger \hat{a}_{1,j}, \\ 
        \hat{L}^{-}_{j,j+1} &= \hat{a}_{1,j}^\dagger \hat{a}_{0,j}, \\
	\hat{L}^{z}_{j,j+1} &= \frac{1}{2}\left(\hat{a}_{0,j}^\dagger \hat{a}_{0,j} - \hat{a}_{1,j}^\dagger \hat{a}_{1,j}\right),
	\end{align}
\end{subequations}
which become the gauge ($\hat{L}^{\pm}_{j,j+1}$) and electric ($\hat{L}^{z}_{j,j+1}$) fields in a QLM formulation. Since the local occupation $\hat{a}_{0,j}^\dagger \hat{a}_{0,j} + \hat{a}_{1,j}^\dagger \hat{a}_{1,j}=2S$ can be high, as large as thousands of atoms \cite{Kasper2017,Mil2020}, this approach permits to work in the large-$S$ limit of the QLM theory. 

The species $a$ is considered sufficiently tightly trapped, such that dynamics can happen only within a site of the optical lattice. 
The essentially only way for it to evolve is then through heteronuclear boson--fermion spin-changing collisions, which preserve the total angular momentum --- thanks to the tight trapping --- locally within each optical-lattice site. 
The corresponding process is described by 
$\hat{H}_\mathrm{s.c.c.} = \Gamma \sum_j \hat{b}_{j,\mathrm{p}}^\dagger \hat{L}^{-}_{j,j+1} \hat{b}_{j+1,\mathrm{v}}+\mathrm{H.c.}$ 
This process encodes the gauge-invariant tunnelling. 

Additional terms come from the local on-site energies $\pm\Delta$ of the internal atomic states $\ket{\mathrm{v}}$ and $\ket{\mathrm{p}}$, as well as on-site Hubbard interactions $\gamma$ between $\ket{\uparrow}$ and $\ket{\downarrow}$ states representing the gauge sector. Together, these realize a gauge theory of the form 
\begin{align}\nonumber
    \hat{H}_{\mathrm{U}(1)} =&\, \hbar\Omega \sum_j( \hat{b}_{j,\mathrm{p}}^\dagger\hat{b}_{j,\mathrm{v}}+\mathrm{H.c.})+ \gamma \sum_j (\hat{L}^{z}_{j,j+1})^2\\\nonumber  
    &+ \Gamma \sum_j (\hat{b}_{j,\mathrm{p}}^\dagger \hat{L}^{-}_{j,j+1} \hat{b}_{j+1,\mathrm{v}}+\mathrm{H.c.})  \\
    &+ \frac{\Delta}{2} \sum_j (\hat{b}_{j,\mathrm{p}}^\dagger\hat{b}_{j,\mathrm{p}}- \hat{b}_{j,\mathrm{v}}^\dagger\hat{b}_{j,\mathrm{v}}). 
\end{align}
This Hamiltonian commutes with the generator of the $\mathrm{U}(1)$ Gauss's law, see Eq.~\eqref{eq:G}, $\hat{G}_j= \hat{L}^{z}_{j,j+1} - \hat{L}^{z}_{j} + \hat{Q}_j$, with $\hat{Q}_j=\hat{b}_{j,\mathrm{p}}^\dagger\hat{b}_{j,\mathrm{p}}+ \hat{b}_{j,\mathrm{v}}^\dagger\hat{b}_{j,\mathrm{v}}$.  
In contrast to other approaches, this layout permits to physically place the gauge fields on the same site as the two matter components it is supposed to interact with, thus greatly enhancing the overlap of the involved fields and consequently the corresponding interaction rate.

This principle was realized for a building block in the experiment of Mil \textit{et al.}~\cite{Mil2020} using a mixture of condensed \textsuperscript{7}Li and \textsuperscript{23}Na atoms. In these experiments, a bosonic species was employed, which permitted to use larger occupation numbers and thus to increase the relevant interaction rates. 

The experiment employed ${300\times 10^3}$ sodium and ${50\times 10^3}$ lithium atoms, and used an external magnetic bias field to energetically suppress any spin change not coming from heteronuclear collisions and to populate only the two Zeeman levels $m_F = 0$ and $1$ of the $F=1$ hyperfine ground-state manifolds. 
This then permitted to observe the gauge-invariant production of (bosonic) particles out of the matter vacuum, purely through interactions with the gauge field (see Fig.~\ref{Fig3}c, bottom). 
Although the implementation was for a single building block, the employed experimental ingredients are in principle all scalable to a large chain. 

\subsection{Chiral BF theory from Raman-dressing}
In the systems reviewed above, gauge and matter fields were directly implemented, which leads to the requirement to enforce the corresponding local gauge symmetries between the constituents. An alternative approach is to focus on theories with either the gauge or matter fields already formally integrated out. As a result, no or fewer local constraints remain. This can alleviate experimental overhead, but at the same time more complex interactions can emerge, which can make direct experimental implementations more challenging. Such a procedure was utilized successfully in a first realization of the $\mathrm{U}(1)$ Schwinger model ($1+1$D lattice QED) on a digital quantum computer using trapped ions~\cite{Martinez2016}.

The same approach has recently been chosen in a cold-atom experiment by Fr\"olian \textit{et al.}~\cite{Froelian2022}, who implemented the so-called chiral BF theory. The latter corresponds to a dimensional reduction of the $2+1$D $\mathrm{U}(1)$ Chern-Simons theory, which is able to describe the gapless chiral edge excitations that appear in fractional quantum Hall systems, see top panel in Fig.~\ref{Fig3}d. Using the local conservation laws to eliminate the gauge degrees of freedom, the chiral BF Hamiltonian becomes
\begin{equation}
    \hat{H}_{\rm chBF} = \int dx \left( - \frac{1}{2m} \hat{\phi}^\dagger \partial_x^2 \hat{\phi} + V(\hat{n}) + \frac{\lambda}{2} \hat{\phi}^\dagger \hat{j} \hat{\phi} \right).
    \label{eqHchBF}
\end{equation}
Here, $V(\hat{n})$ describes a local interaction of the matter field $\hat{\phi}(x)$ (polynomial in $\hat{n}(x) = \hat{\phi}^\dagger(x) \hat{\phi}(x)$), and the last term constitutes the chiral BF interaction which involves the current operator $\hat{j}(x) = [ \hat{\phi}^\dagger \partial_x \hat{\phi} - (\partial_x \hat{\phi}^\dagger) \hat{\phi}] / (2 i m)$ and breaks Galilean invariance.

The Hamiltonian~\eqref{eqHchBF} was implemented by Fr\"olian \textit{et al.}~\cite{Froelian2022} in a Raman-dressed Bose-Einstein condensate (BEC). The obtained momentum-dependent spin texture leads to a momentum-dependent interaction, as required for the chiral BF theory. The experimentalists demonstrated their method by realizing a chiral soliton in their setup, which corresponds to a nonlinear nondispersing solution of the corresponding wave equation. Notably, the soliton becomes unstable when its momentum is reversed: This is reflected by the momentum-dependent width of the wave packet observed over time, see bottom panel of Fig.~\ref{Fig3}d, where the blue dots (green squares) correspond to the stable chiral soliton (unstable wave packet) with $k_x<0$ ($k_x>0$).

The chiral BF theory can be viewed as a point of departure for realizing a larger class of two-dimensional topological field theories with anyon excitations captured by Chern-Simons gauge theory~\cite{ValentiRojas2020}.

\section{Current state of the art: Large-scale gauge-theory cold-atom quantum simulators}\label{sec:LargeScaleCAQS}

We now turn our attention to the current experimental state-of-the-art of cold-atom quantum simulators of gauge theories, which comprise some of the largest-scale platforms for probing gauge-theory dynamics from first principles. We review two seminal experiments, as summarized in Fig.~\ref{fig:mapping}.

\subsection{Rydberg setups}
The experiments of Bernien \textit{et al.}\ \cite{Bernien2017} realized a one-dimensional tweezer array (length $L$) with a variable number of Rydberg excitations created by laser light. 
The electronic ground and Rydberg states form a pseudo-spin-$1/2$, with states $\Ket{\downarrow}_j$ and $\Ket{\uparrow}_j$, and associated Pauli operators $\hat\tau^\alpha_j$. Here, $j$ labels the tweezer trap. 
The corresponding dynamics is described by the Hamiltonian
 \begin{equation}     
 \label{eq:HRydberg}
 \hat H_{\text{Ryd}} = \sum_{j=1}^{L}(\Omega \,\hat\tau^x_j +\delta\, \hat\tau^z_j) + \sum_{j < \ell=1}^{L} V_{j,\ell} \hat n_j\hat n_\ell.
  \end{equation}
The operator $\hat n_j = (\hat \tau^z_j+1)/2$ counts the atoms in the Rydberg state, which interact with strength $V_{j,\ell}$. Further, $2\Omega$ and $2\delta$ are the Rabi frequency and the detuning of the laser excitation. 
The experiments of Ref.~\cite{Bernien2017} focused on the regime where $V_{j,j+1}$ is much larger than all other energy scales (due to the sharp decay of Rydberg interactions with distance, $V_{j,\ell}$ beyond nearest neighbors can be neglected in what follows). 
The resulting Rybderg blockade leads to the constraint $\hat n_j\hat n_{j+1} = 0$, i.e., atoms on neighboring sites cannot simultaneously reside in the Rydberg state. 
The Fendley--Sengupta--Sachdev (FSS) Hamiltonian describing this regime reads \cite{Fendley2004}  
 \begin{equation}
 \label{eq:Hfss}
 \hat H_{\text{FSS}} = \mathfrak{P}_{\mathrm{Rb}}\sum_{j=1}^L \left(\Omega \, \hat \tau_j^x + \delta\, \hat \tau_j^z\right)\mathfrak{P}_{\mathrm{Rb}}.
 \end{equation}
Despite its simple-looking form, Rydberg excitations become strongly correlated across different traps owing to the applied projector $\mathfrak{P}_{\mathrm{Rb}}$ onto the subspace respecting the Rydberg blockade, making the FSS Hamiltonian an instance of the famous PXP model~\cite{Turner2018}.

After the experimental study~\cite{Bernien2017} was published, Surace \textit{et al.}~\cite{Surace2020} pointed out that the FSS Hamiltonian can be mapped to the spin-$1/2$ $\mathrm{U}(1)$ QLM by using Gauss's law to integrate out the matter fields. This implies that the experimental results by Bernien \textit{et al.}~\cite{Bernien2017} can be interpreted in the context of LGTs, which is the perspective we will take in the following.

To understand how the gauge theory arises, consider a ($1+1$D) QLM with arbitrary spin-$S$ and target sector $g_\ell=0,\,\forall\ell$. The Gauss's law, Eq.~\eqref{eq:G}, allows to eliminate the matter field by writing $\hat{\sigma}^z_\ell=-2(\hat{s}^z_{\ell-1,\ell}+\hat{s}^z_{\ell,\ell+1})-1$. Inserting this relation into the Hamiltonian of the QLM, Eq.~\eqref{eq:QLM}, the latter can be rewritten as
\begin{align}
\label{eq:HPXP}
    \hat{H}_\text{spin}=&\,\hat{\mathfrak{P}}\sum_\ell\bigg[-\frac{\kappa}{a\sqrt{S(S+1)}}\hat{s}^x_{\ell,\ell+1}-2\mu\hat{s}^z_{\ell,\ell+1}\\\label{eq:PXP}
    &+\frac{g^2a}{2}\big(\hat{s}^z_{\ell,\ell+1}\big)^2+a\chi(-1)^{\ell+1}\hat{s}^z_{\ell,\ell+1}\bigg]\hat{\mathfrak{P}},
\end{align}
where $\hat{\mathfrak{P}}$ is the projector onto gauge configurations consistent with the target sector $g_\ell=0,\,\forall\ell$. 
For the case of $S=1/2$ relevant to the Rydberg experiment~\cite{Bernien2017}, the term $\propto g^2a/2$ becomes a constant; hence, by identifying $\hat{\tau}_j=\hat{s}_{\ell-1,\ell}$, $\Omega=-\kappa/[a\sqrt{S(S+1)}]$, and $\delta=-2\mu$, the above Hamiltonian for $\theta=\pi$, i.e., $\chi=0$, coincides with $\hat{H}_{\text{FSS}}$. 

\begin{figure}
	\centering	\includegraphics[width=\linewidth]{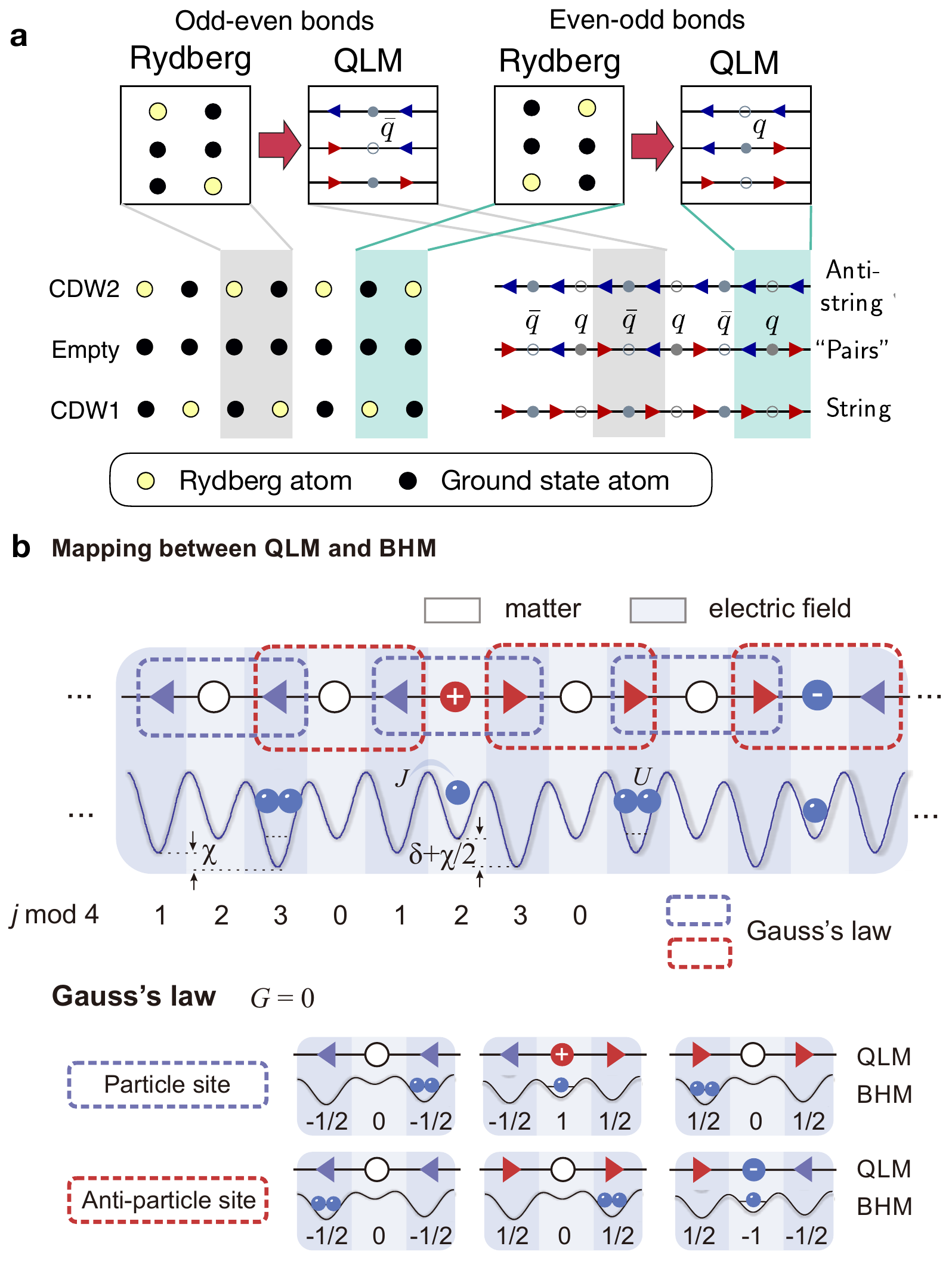}	

\caption{\textbf{Large-scale realizations of LGTs with cold atoms.} (a) Mapping by Surace et al.~\cite{Surace2020} of the FSS Hamiltonian, describing an array of strongly interacting and driven Rydberg atoms, to a $\mathrm{U}(1)$ QLM. The Rydberg-blockaded state with two adjacent Rydberg excitations corresponds to the gauge-field configuration forbidden after integrating out the matter excitations. The resulting QLM describes the Rydberg-tweezer experiment by Bernien et al.~\cite{Bernien2017}. 
(b) Mapping of the spin-$1/2$ $\mathrm{U}(1)$ QLM~\eqref{eq:QLM} onto an optical superlattice of cold bosons governed by the microscopic Bose--Hubbard Hamiltonian~\eqref{eq:BHM}. Even (odd) sites of the optical lattice represent the sites (links) of the QLM, where a staggering potential $\delta$ distinguishes between odd and even sites on the optical lattice. An even site can host a single boson, indicating the presence of charged matter, or no bosons at all, indicating the absence of charged matter. An odd site can host either $0$ or $2$ bosons, representing the two possible polarizations of the local electric field. A second staggering potential $\chi$ is applied onto the odd sites (links of the QLM) in order to realize the topological $\theta$-term.}	\label{fig:mapping}
\end{figure}

Notably, the Rydberg blockade constrains the theory exactly to the physical subspace of states respecting Gauss's law~\cite{Surace2020}: Namely, the $\mathrm{U}(1)$ electric field can only remain unchanged in the case with no charge, or increase (decrease) in the case of a particle (antiparticle) occupying an even (odd) matter site, see Fig.~\ref{fig:mapping}a. This leaves one disallowed gauge-field configuration per matter site, which maps onto the pair of neighboring Rydberg excitations avoided by the blockade mechanism. Finally, Surace \textit{et al.}~\cite{Surace2020} also pointed out how a tunable $\theta$-angle can be added to the Rydberg setup, e.g., by a position-dependent AC Stark shift.  

The measurements by Bernien et al.~\cite{Bernien2017} started from a charge density wave (CDW) state of alternating Rydberg-no Rydberg excitations along the chain and observed the subsequent dynamics. Remarkably, they found slow and only weakly damped oscillations to the reversed CDW state and back. Initially, these oscillations were interpreted as quantum scars \cite{Turner2018}. Using the mapping to a $\mathrm{U}(1)$ QLM, Surace \textit{et al.}~\cite{Surace2020} were able to relate these oscillations directly to string-inversion dynamics well-known to occur in various gauge theories, clarifying the origin of this phenomenon more broadly.

\subsection{Gauge protection}
\label{sec:gaugeprotection}
The principal property of a gauge theory is gauge symmetry. As such, a necessary requirement for a quantum simulator of a gauge theory is to realize a stabilized and controlled gauge symmetry. This requirement becomes particularly important in large-scale quantum simulators, where an extensive number of local constraints have to be controlled. This is indeed a nontrivial problem, as a quantum simulator will generally always have unavoidable gauge symmetry-breaking terms. These will lead to a steady and uncontrolled buildup of gauge violations, eventually rendering the dynamics obtained from the quantum simulator unrepresentative of the underlying gauge theory~\cite{Halimeh2020a}.

A theoretically straightforward way to suppress gauge violations on a quantum simulator involves realizing a ground-state manifold of the target gauge sector, as proposed by Halimeh and Hauke~\cite{Halimeh2020a}. Let us denote the target gauge superselection sector by its background charges $g_j^\text{tar}$. Engineering the protection term $V\hat{H}_G=V\sum_j\big(\hat{G}_j-g_j^\text{tar}\big)^2$ into the quantum simulator will then ensure, at sufficiently large $V$, that the target sector is a ground-state manifold. As a result, processes due to $\lambda\hat{H}_1$, which take the system to other gauge sectors, are energetically penalized, and the gauge violation, defined as $\sum_j\langle\hat{G}_j^2\rangle/L$, settles into a value $\propto\lambda^2/V^2$ up to all numerically accessible times for sufficiently large $V$~\cite{Halimeh2020a}. Although a very effective protection scheme, this approach can require a significant engineering overhead, which can make it experimentally unfeasible.

For the case of a $\mathrm{U}(1)$ LGT, there is a much simpler way to protect gauge symmetry by employing the method of \textit{linear gauge protection}, proposed by Halimeh \textit{et al.}~\cite{Halimeh2020e}, where the following protection term is employed: $V\hat{H}_G=V\sum_jc_j\hat{G}_j$. If the coefficients $c_j$ are tailored to satisfy the \textit{compliance} condition $\sum_jc_jg_j^\text{tar}\neq\sum_jc_jg_j$ where at least one $g_j\neq g_j^\text{tar}$, then stabilization of the gauge symmetry is guaranteed up to times exponential in a volume-independent $V$, with the gauge violation again settling into a plateau of value $\propto\lambda^2/V^2$. However, this becomes untenable for large systems, as then the coefficients $c_j$ will grow exponentially, leading to experimental impracticability. Fortunately, it turns out that the compliance condition is not necessary for stabilizing the gauge symmetry for most experimentally relevant \textit{local} errors. Indeed, one can analytically show that when the coefficients are simply $c_j=(-1)^j$, the gauge violation is once again suppressed into a plateau $\propto\lambda^2/V^2$ up to times linear $V$ in a worst-case scenario. Even more, numerical benchmarks using exact diagonalization~\cite{Halimeh2020e} and matrix product states~\cite{vandamme2021reliability} show that the stabilization of the gauge symmetry can persist up to all accessible timescales independent of system size. As we will discuss in the following, this method of linear gauge protection has proven crucial in stabilizing the gauge symmetry in the first large-scale cold-atom quantum simulators of gauge theories.

\twocolumngrid
\definecolor{shadecolor}{rgb}{0.8,0.8,0.8}
\begin{shaded}\label{box:mapping}
\noindent{\bf Box 2 $|$ Bosonic mapping}
\end{shaded}
\vspace{-9mm}
\definecolor{shadecolor}{rgb}{0.9,0.9,0.9}
\begin{shaded}
\noindent
In the spin-$1/2$ representation, the electric-field operator $\hat{s}^z_{\ell,\ell+1}$ is a two-level system with eigenvalues $\pm1/2$. This automatically means that the gauge-coupling term in Hamiltonian~\eqref{eq:QLM} is rendered an inconsequential constant, which we can neglect. The matter-field (Pauli) operator $\hat{\sigma}^z_\ell$ is also a two-level system, regardless of $S$. Let us now imagine an optical superlattice whose even (odd) sites shall represent the sites (links) of the $\mathrm{U}(1)$ QLM, see Fig.~\ref{fig:mapping}b. If we denote the sites of this optical superlattice with the index $j$, then we have the correspondence: even sites $j(\ell)=2\ell$ and odd sites $j(\ell,\ell+1)=2\ell+1$, with the latter corresponding to the link between the QLM sites $\ell$ and $\ell+1$. On the even sites of the optical lattice, a hard-core boson constraint should be enforced to faithfully represent the matter field: 
\begin{subequations}
    \begin{align}
        \hat{\sigma}^+_{\ell}&=\hat{\mathcal{P}}_\ell\hat{a}_\ell^\dagger\hat{\mathcal{P}}_\ell,\\
        \hat{\sigma}^z_{\ell}&=\hat{\mathcal{P}}_\ell(2\hat{a}_\ell^\dagger\hat{a}_\ell-1)\hat{\mathcal{P}}_\ell,
    \end{align}
\end{subequations}
where $\hat{\mathcal{P}}_\ell$ is the projector onto the local bosonic Hilbert space $\text{span}\{\ket{0}_\ell,\ket{1}_\ell\}$. On the odd sites of the superlattice, which host the electric and gauge fields of the QLM, a doublon constraint is enforced: 
\begin{subequations}
    \begin{align}
        \hat{s}^+_{\ell,\ell+1}&=\frac{1}{\sqrt{2}}\hat{\mathcal{P}}_{\ell,\ell+1}\hat{a}_{\ell,\ell+1}^\dagger\hat{a}_{\ell,\ell+1}^\dagger\hat{\mathcal{P}}_{\ell,\ell+1},\\
        \hat{s}^z_{\ell,\ell+1}&=\frac{1}{2}\hat{\mathcal{P}}_{\ell,\ell+1}(\hat{a}_{\ell,\ell+1}^\dagger\hat{a}_{\ell,\ell+1}-1)\hat{\mathcal{P}}_{\ell,\ell+1},
    \end{align}
\end{subequations}
where $\hat{\mathcal{P}}_{\ell,\ell+1}$ is the projector onto the local bosonic Hilbert space $\text{span}\{\ket{0}_{\ell,\ell+1},\ket{2}_{\ell,\ell+1}\}$. The bosonic ladder operators satisfy the canonical commutation relations $[\hat{a}_j,\hat{a}_i]=0$ and $[\hat{a}_j,\hat{a}_i^\dagger]=\delta_{j,i}$.

\end{shaded}

\twocolumngrid

Finally, it is worth noting that linear gauge protection has been extended to other Abelian gauge theories. For example, in the case of the $\mathbb{Z}_2$ LGTs, the concept of a \textit{local pseudogenerator} $\hat{W}_j$ has been introduced. It acts identically to the full local generator $\hat{G}_j$ in the target sector, but is actually a two-body instead of a three-body term, easing experimental requirements~\cite{Halimeh2022stabilizing,Homeier2023}. One then modifies the protection term to $V\hat{H}_W=V\sum_jc_j\hat{W}_j$, and the qualitative picture of gauge-symmetry stabilization remains the same as above.

\begin{figure*}
	\centering	\includegraphics[width=0.9\linewidth]{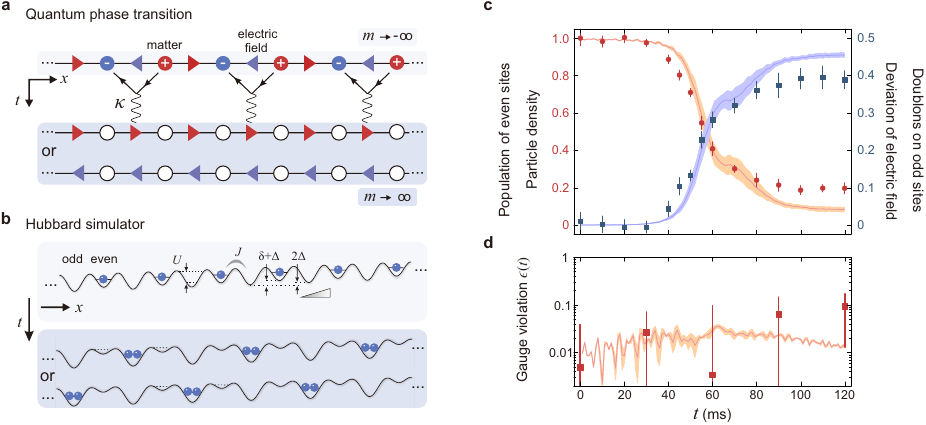}	

\caption{\textbf{Simulation of a lattice gauge theory in a $71$-site quantum system.} In Ref.~\cite{Yang2020}, Yang \emph{et al.} experimentally observed the quantum phase transition and the fundamental gauge symmetry of a $1+1$D $\mathrm{U}(1)$ LGT. (a) The Feynman diagram depicts the quantum phase transition from a charge-proliferated phase (with particle rest mass $m \rightarrow -\infty$) to a vacuum phase (with $m \rightarrow +\infty$), illustrating the gauge-invariant annihilation of particles and antiparticles. In the final vacuum state, a charge-conjugation--parity symmetry breaking phase emerges, allowing the electric field to freely traverse the system in two opposite directions. (b) The matter and gauge field of the LGT are mapped onto the occupation number of ultracold atoms in a superlattice.
In this $71$-site Hubbard simulator, the on-site interaction is denoted as $U$, tunneling strength as $J$ and the energy offsets as $\delta$ and $\Delta$. The state where each even site contains one atom corresponds to the charge-proliferated state with $m \rightarrow -\infty$. Conversely, in the limit of $m \rightarrow +\infty$, the ground state is a vacuum phase where doublons occupy half of the odd lattice sites. The vacuum phase manifests two distinct configurations, each corresponding to one direction of the electric field as shown in (a).
(c) The quantum phase transition occurs when the rest mass $m$ is adiabatically ramped from a large negative to a large positive value.
The transition leads to the transfer of atom population from even sites to odd sites, resulting in the formation of doublons. The deviation of the electric field serves as the order parameter in this phase transition. (d)
The probabilities of the gauge-invariant states are measured during the many-body dynamics. The numerical results are shown in the orange curve. Throughout the phase transition, the gauge violation, quantified by Gauss's law, remains below $10\%$, indicating the preservation of gauge invariance.
}\label{Fig6}
\end{figure*}

\subsection{Mapping onto optical superlattice}

\subsubsection{With matter fields}
We are now ready to discuss large-scale cold-atom quantum simulators of gauge theories, including how the method of linear gauge protection stabilizes their gauge symmetry and allows large system sizes. Our focus will be on the spin-$1/2$ $\mathrm{U}(1)$ QLM, i.e., Eq.~\eqref{eq:QLM} with $S=1/2$, which we map onto a cold-atom quantum simulator.

Starting with Hamiltonian~\eqref{eq:QLM} and employing the bosonic mapping outlined in Box 2, we arrive at the effective Hamiltonian

\begin{align}\nonumber
    \hat{H}=\hat{\mathcal{P}}\sum_\ell\bigg\{&-\frac{\kappa}{\sqrt{6}}\Big[\hat{a}_\ell\big(\hat{a}^\dagger_{\ell,\ell+1}\big)^2\hat{a}_{\ell+1}+\text{H.c.}\Big]\\\label{eq:Heff}
    &+\mu\hat{a}_\ell^\dagger\hat{a}_\ell+\frac{\chi}{2}(-1)^{\ell+1}\hat{a}_{\ell,\ell+1}^\dagger\hat{a}_{\ell,\ell+1}\bigg\}\hat{\mathcal{P}},
\end{align}
where $\hat{\mathcal{P}}=\prod_\ell\hat{\mathcal{P}}_\ell\hat{\mathcal{P}}_{\ell,\ell+1}$ is a projector onto the target sector of Gauss's law.

In order to realize Hamiltonian~\eqref{eq:Heff} in an actual cold-atom setup, we need to consider the Bose--Hubbard Hamiltonian on an optical superlattice, which naturally governs the microscopic dynamics of the constituent bosons, and constrain it into an excited manifold that leads to the effective Hamiltonian~\eqref{eq:Heff}. The Bose--Hubbard Hamiltonian is given by

\begin{align}\nonumber
    \hat{H}_\text{BH}=&-J\sum_{j=1}^{L-1}\big(\hat{a}_j^\dagger\hat{a}_{j+1}+\text{H.c.}\big)+\frac{U}{2}\sum_{j=1}^L\hat{n}_j\big(\hat{n}_j-1\big)\\\label{eq:BHM}
    &+\frac{1}{2}\sum_{j=1}^L\big[(-1)^j\delta+2j\Delta+\chi_j\big]\hat{n}_j,
\end{align}
where $J$ is the tunneling strength, $U$ is the on-site repulsion potential, $\delta$ is the staggering potential distinguishing between matter sites (even $j$) and gauge links (odd $j$), $\Delta$ is a linear tilt, $L=2L_\text{m}$ is the number of sites, and $\hat{n}_j=\hat{a}_j^\dagger\hat{a}_j$. Furthermore, we have added a second staggering potential on odd sites such that $\chi_j=\pm\chi$ when $j\mod4=1$ and $3$, respectively, ($\chi_j=0$ otherwise) allowing us to realize a topological $\theta$-term at finite $\chi$. In the regime $U\approx2\delta\gg\Delta,J,\chi>0$, we are able to restrict the allowed bosonic occupations on odd (even) sites of the optical superlattice to $0$ and $2$ ($0$ and $1$), in accordance with our desired mapping between the local degrees of freedom of the QLM and those of the bosonic model. This then allows us to obtain the effective Hamiltonian~\eqref{eq:Heff} from the microscopic Bose--Hubbard Hamiltonian~\eqref{eq:BHM} up to second order, $\mathcal{O}(J^2/U)$, in degenerate perturbation theory, where
\begin{subequations}
\begin{align}
    \kappa&\approx\sqrt{6}J^2\bigg[\frac{\delta}{\delta^2-\Delta^2}+\frac{U-\delta}{(U-\delta)^2-\Delta^2}\bigg],\\
    \mu&\approx\delta-\frac{U}{2}.
\end{align}
\end{subequations}

Examining Hamiltonian~\eqref{eq:BHM}, one can see that the tunneling term perturbatively breaks gauge symmetry, but is essential to induce gauge-theory dynamics. The gauge-invariant part of this Hamiltonian can be written as
\begin{align}\nonumber
    \hat{H}_\text{diag}=\sum_\ell\bigg\{&\frac{U}{2}\big[\hat{n}_\ell\big(\hat{n}_\ell-1\big)+\hat{n}_{\ell,\ell+1}\big(\hat{n}_{\ell,\ell+1}-2\big)\big]\\
    &+\frac{1}{2}\big[(-1)^\ell\chi-2\mu\big]\hat{n}_{\ell,\ell+1}+c_\ell\hat{G}_\ell\bigg\},
\end{align}
where the linear gauge protection term $\sum_\ell c_\ell\hat{G}_\ell$ emerges with $c_\ell=2(-1)^\ell\ell\Delta$, and the generator~\eqref{eq:G} is rewritten in terms of the bosonic operators as

\begin{align}
\hat{G}_\ell=(-1)^\ell\bigg(\hat{n}_\ell+\frac{\hat{n}_{\ell-1,\ell}+\hat{n}_{\ell,\ell+1}}{2}-1\bigg).
\end{align}
This linear gauge protection term is what stabilizes the gauge symmetry of this quantum simulator, and the tilt potential $\Delta$ is crucial for its emergence. From the perspective of degenerate perturbation theory, the tilt potential $\Delta$ suppresses gauge symmetry-breaking second-order hopping processes to next-to-nearest-neighbor sites so long as $\Delta\gg J^2/\delta$.

\subsubsection{Integrating out the matter fields}\label{sec:IntegrateOut}

It is also possible to realize a gauge theory in an optical lattice by integrating out the matter fields. 
Namely, the PXP-Hamiltonian from Eq.~\eqref{eq:HPXP} for spin-$1/2$ maps onto the Bose--Hubbard model \cite{Su2023}
\begin{align}\nonumber
    \hat{H}'_\text{BH}=&-J\sum_{j=1}^{L-1}\big(\hat{a}_j^\dagger\hat{a}_{j+1}+\text{H.c.}\big)+\frac{U}{2}\sum_{j=1}^L\hat{n}_j\big(\hat{n}_j-1\big)\\\label{eq:BHMp}
    &+\frac{1}{2}\sum_{j=1}^L\big[(-1)^j\chi+2j\Delta\big]\hat{n}_j.
\end{align}
Note that here there is no $\delta$, and this makes sense as there are only links in Hamiltonian~\eqref{eq:PXP}, obviating the need of a staggering potential to differentiate them from the nonexistent matter sites. Furthermore, this mapping works in the regime of $U\approx\Delta\gg\chi,J$, where then $\mu\approx(\Delta-U)/2$ and $\kappa\approx\sqrt{6}J$.

\subsection{Experiments in optical lattices}
\label{sec:largescaleU1}
The large-scale lattice gauge theory (LGT) in the Hubbard model was successfully quantum-simulated in a recent experiment conducted by Yang \emph{et al.}~\cite{Yang2020}.
This achievement was made possible by state-of-the-art advancements in cold-atom manipulation techniques. The preservation of the gauge invariant subspace was first of all ensured through a high-fidelity state preparation procedure. Starting from a quasi $2$D Bose-Einstein condensate, the atoms were further cooled in an optical superlattice to create a nearly unity occupied atomic MI~\cite{Yang2020Cooling}. The filling factor of this Mott state was measured to be 0.992(1) in a $2$D square lattice consisting of ten thousand sites. 

To effectively separate the $2$D MI into an array of $1$D systems, the coupling along one direction was suppressed. This allowed for the realization of a large-scale quantum simulation in a $1$D system. Notably, in such a $1$D system with 100 lattice sites, the average length of a defect-free chain was approximately 71 sites. The employed optical superlattice further divided the atoms into even and odd sites, representing the matter and gauge sites, respectively. In the state initialization, a sub-lattice spin addressing technique is employed to remove atoms residing on the even sites~\cite{Yang2020Cooling}. The presence of both the matter field and gauge field was a crucial element for observing gauge invariance.

The experiment demonstrated complete control over the model parameters to manipulate Coleman's phase transition~\cite{Coleman1976}, see Fig.~\ref{Fig6}a. In addition to regulating on-site interaction $U$ and tunneling term $J$, the superlattice potential employed two types of energy offsets ($\delta$ and $\Delta$) to finely tune the model parameters, as sketched in Fig.~\ref{Fig6}b. The ground state at the extreme negative rest mass $m \rightarrow -\infty$ was characterized by the presence of a single atom on the matter (even) sites, while the gauge (odd) sites remain unoccupied (see Fig.~\ref{Fig6}a). The quantum phase transition was driven by slowly sweeping the rest mass $m$ and the coupling strength $\kappa$ of the system. The ramp speed was optimized to minimize non-adiabatic excitations and unwanted heating effects. As the rest mass approached large positive values shown in Fig.~\ref{Fig6}a,b, the system converged to the ground state of another limit with $m \rightarrow +\infty$. In this state, a majority of atoms were transferred from even to odd sites, forming doublons and corresponding to the annihilation of particle and anti-particle pairs.

The phase transition process was observed through time-dependent measurements of site occupation, as depicted in Fig.~\ref{Fig6}c. Following the phase transition, the development of spatial order was quantified using the density-density correlation method. A significant contribution of this experiment is the direct demonstration of the fundamental gauge symmetry. This was achieved by detecting the probabilities of gauge-invariant states throughout the phase transition, allowing for the quantification of violation of Gauss's law or gauge invariance. Despite the imaging system's limitation in achieving single lattice site resolution, the precise state engineering technique was employed to probe the three gauge-allowed Fock states. The degree of gauge violation, denoted as $\epsilon(t)$, represents the projection of the system state outside the gauge-invariant subspace. A upper bound for the gauge violation is presented in Fig.~\ref{Fig6}d.

Quantum simulation holds the potential to outperform classical computers in specific tasks, particularly in the far-from-equilibrium dynamics of gauge theories that involve exponentially large Hilbert spaces. On the experimental platform reviewed above, the thermalization dynamics of a $1+1$D LGT were experimentally investigated using this large-scale Hubbard simulator in Zhou \textit{et al.}~\cite{Zhou2022}. To enforce gauge symmetry and constraints, energy penalties were implemented in an optical superlattice. By controlling the degree of gauge violation, a controllable method was established to distinguish between the gauge invariant and non-constrained regimes.

Upon subjecting the system to a global quantum quench, the unitary dynamics, governed by $\mathrm{U}(1)$ symmetry, exhibited emergent irreversible thermalization behavior. The thermalization dynamics with and without gauge constraints displayed distinct behaviors, particularly in the final steady state. In the gauge constrained subspace, the out-of-equilibrium dynamics demonstrated an effective loss of initial state information, characteristic behavior of a thermal state. By manipulating the energy density of the initial state prior to the quench dynamics, the final states converged to a steady state with the same effective temperature. The equilibration to thermal equilibrium value was observed in  all of the quench dynamics. This research paves the way for exploring far-from-equilibrium dynamics of higher-dimensional gauge theories, where classical computational methods face challenges.

By utilizing a quantum gas microscope, the critical point of Coleman's phase transition was accurately determined in the Hubbard simulator~\cite{Wang2023}. The investigation focused on the equilibrium and quench dynamics in the vicinity the critical regime. The critical point was located through finite-size scaling analysis, where the intersection point of the phase transition curves was identified. Additionally, the results revealed that the phase transition exhibited a universal scaling of Ising type.

Su \textit{et al.}~\cite{Su2023} adapted this setup by integrating out the matter fields, as described in Sec.~\ref{sec:IntegrateOut}, and uncovered rich scarring regimes. The experiment employed quantum-interference protocols to measure entanglement entropy, which showed that upon preparing the system in a scarred state, the subsequent many-body dynamics is trapped in a low-entropy subspace. This simplified version of the optical lattice, given in Eq.~\eqref{eq:BHMp}, opens the door for manipulating $\chi$ with a staggered superlattice potential and subsequently tuning the topological $\theta$-term.

The confinement--deconfinement phase transition can be observed by tuning the topological $\theta$-term in Eq.~\eqref{eq:QLM}. Based on theoretical proposals by Halimeh \textit{et al.}~\cite{Halimeh2022tuning} and Cheng \textit{et al.}~\cite{Cheng2022}, a recent experiment by Zhang \textit{et al.}~\cite{Zhang2023} studied the confinement--deconfinement transition by manipulating $\chi$ with a staggered lattice potential; see Eq.~\eqref{eq:BHMp}. Within this setup, a particle-antiparticle pair was prepared within the $1$D vacuum state background of the QLM. Real-time dynamics were monitored during a large-mass quench at various value of $\chi$. At small $\chi$, the length of electric string between the particle-antiparticle pairs grew ballistically. Conversely, at larger values of $\chi$, the string length remained close to its initial value, indicating the emergence of the confinement phase. This allowed for a clear distinction between the confined and deconfined phases.

\section{Perspective}\label{sec:perspective}
We have reviewed the exciting experimental progress in the quantum simulation of LGTs using the cold-atom platform. In order to be able to probe gauge-theoretic phenomena of relevance to experiments such as at the LHC and RHIC, this technology must be further advanced, in particular to accommodate higher spatial dimensions \cite{Zohar_NewReview}, larger representations of the electric field, and non-Abelian gauge groups. Below, we outline a few proposals towards overcoming these main challenges.

{\emph{Abelian: higher dimensions, higher-spin representations.---}}
Various proposals have been put forward to further advance cold-atom quantum simulators of Abelian gauge groups. For example, various mappings of the electric-field operator onto bosonic occupation numbers have been suggested to achieve higher-level representations of the electric field \cite{Banerjee2012,Yang2016,Zohar2013,Ott2020scalable}. 
A higher-spin mapping leveraging on the scheme discussed in Sec.~\ref{sec:largescaleU1} has recently been presented in Ref.~\cite{osborne2023spins}. There, the local electric-field eigenvalue $s^z_{\ell,\ell+1}$ is represented by the bosonic occupation $n_{\ell,\ell+1}=2(s^z_{\ell,\ell+1}+S)$, which converges faithfully to the lattice-QED limit for $S\to\infty$, and an extended Bose--Hubbard setup with bosonic Dysprosium atoms has been proposed for a three-level representation of the electric field. This extension allows for the observation of string breaking \cite{Banerjee2012} as well as for a tunable gauge-coupling term that is not possible in the two-level representation. Such a term is a crucial parameter for confinement in QED, and its realization on a quantum simulator would therefore be a significant step forward. 

A $2+1$D generalization of this quantum simulator has also been proposed to realize a quantum link formulation of scalar QED \cite{osborne2022largescale}, where the matter degrees of freedom are hard-core bosons. The gauge protection scheme in this case involved a tilt in both spatial dimensions, but was shown to also be very effective in numerical benchmarks. 
Other proposals to advance cold-atom gauge quantum simulators to higher dimensions include approaches based on Floquet engineering \cite{Dutta2017}, digital optical-lattice schemes \cite{Zohar2017}, $\mathbb{Z}_2$ gauge protection in Rydberg tweezer setups \cite{Homeier2023}, as well as the exploitation of interspecies spin-changing collisions between small condensates located at the links of an optical lattice~\cite{Ott2020scalable}, as a variation of the setup discussed in Sec.~\ref{sec:U1BuildingBlock}.

{\emph{Non-Abelian in $d+1$D.---}}
The importance of proceeding to non-Abelian gauge symmetries in view of applications to high-energy and nuclear physics questions has been realized early on. Pioneering proposals included exploitation of Rydberg interactions \cite{Tagliacozzo2013}, symmetries of the interactions of cold alkaline-earth atoms in species-dependent optical superlattices \cite{Banerjee2013}, and angular-momentum conservation in interacting boson--fermion mixtures \cite{Zohar2013} (similar in spirit to the simplified Abelian approach of Sec.~\ref{sec:U1BuildingBlock}). 
Many further proposals have since been presented on the cold-atom platform, with recent years having seen a stronger focus on digital approaches such as in Rydberg arrays \cite{Gonzalez2022}. 

A recent proposal \cite{Halimeh2023Spin} employed a \textit{top-down} approach making use of gauge protection in the spirit of Sec.~\ref{sec:gaugeprotection}. In that proposal, the emphasis was on realizing an $\mathrm{SU}(N)$ gauge symmetry locally at vertices, while obtaining an effective LGT by perturbatively inducing tunneling between neighboring vertices. For example, by employing cold polar molecules in optical tweezer arrays, one can utilize the naturally occurring dipole interactions between these molecules at a vertex to realize an $\mathrm{SU}(2)$ gauge symmetry through a proper Floquet sequence. Extensions of this approach to $\mathrm{SU}(N)$ Hubbard models have also been proposed \cite{Halimeh2023Spin,Surace2023scalable}.

Despite much progress on the side of proposals, the cold-atom quantum simulation of non-Abelian symmetries still remains an outstanding challenge for experiments, since it involves the precise control of a significantly larger number of degrees of freedom (such as the color charge of QCD).

In conclusion, throughout the last decade or so, the cold-atom quantum simulation of gauge theories has been an extremely lively area of research. While of much potential already by itself, it has also engendered stimulating cross-fertilization with various other fields:  
extremely promising proposals and groundbreaking implementations on other platforms have been demonstrated, most notably superconducting qubits and trapped ions \cite{Martinez2016,Klco2018,Kokail2019,Wang2021,Mildenberger2022,Huffman2022,Mueller2023}, 
quantum simulation of gauge theories presents stimulating connections between subatomic physics and condensed-matter models such as spin ice and topological states of matter \cite{Moessner2001,Balents2002,Hermele2004,Glaetzle2014} as well as topological quantum technologies \cite{Kitaev2003,Lechner2015}; 
and quantum simulation and classical simulation methods, such as tensor networks or path-integral approaches, are driving each other \cite{Berges_review}.  
As these examples and those treated in the bulk of this article illustrate, though much theoretical, experimental, and engineering work still lies ahead on the road, the cold-atom quantum simulation of LGTs is an exciting arena of current research.

\section*{Acknowledgments} 
J.C.H.~and F.G.~acknowledge funding from the European Research Council (ERC) under the European Union’s Horizon 2020 research and innovation programm (Grant Agreement no 948141) — ERC Starting Grant SimUcQuam.
J.C.H., M.A., and F.G.~acknowledge funding from the Deutsche Forschungsgemeinschaft (DFG, German Research Foundation) under Germany's Excellence Strategy -- EXC-2111 -- 390814868 and from the European Union’s Horizon 2020 research and innovation programme under Grant Agreement No 101017733 (DYNAMITE, DFG project number 499183856).
F.G.~and M.A.~acknowledge funding from the Deutsche Forschungsgemeinschaft (DFG, German Research Foundation) via Research Unit FOR 2414 under project number 277974659.
M.A.~further acknowledges funding from the ERC under the European Union Horizon 2020 research and innovation program (Grant Agreement No.~803047, LaGaTYb), funding under the Horizon Europe programme HORIZON-CL4-2022-QUANTUM-02-SGA via the project 101113690 (PASQuanS2.1) and funding from the Deutsche Forschungsgemeinschaft (DFG, German Research Foundation) via Research Unit FOR 5522 under project number 499180199.
P.H.~acknowledges support by the ERC Starting Grant StrEnQTh (project ID 804305), Provincia Autonoma di Trento, Q@TN, the joint lab between University of Trento, FBK, INFN, and CNR, and from ICSC - Centro Nazionale di Ricerca in HPC, Big Data and Quantum Computing, funded by the European Union under NextGenerationEU.
The project is funded within the QuantERA II Programme that has received funding from the European Union’s Horizon 2020 research and innovation programme under Grant Agreement No 101017733.
This project has received funding by the European Union under Horizon Europe Programme - Grant Agreement 101080086 - NeQST.
This project has received funding from the Italian Ministry of University and Research (MUR) through the FARE grant for the project DAVNE (Grant R20PEX7Y3A).
Views and opinions expressed are however those of the author(s) only and do not necessarily reflect those of the European Union or the European Commission.
Neither the European Union nor the granting authority can be held responsible for them.
B.Y.~acknowledges support from National Natural Science Foundation of China (Grant No. 12274199), National Key R\&D Program of China (Grant No. 2022YFA1405800) and the Stable Support Plan Program of Shenzhen Natural Science Fund (Grant No. 20220815092422002). 

We thank D.~Banerjee, L.~Barbiero, J.~Berges, A.~Bohrdt, U.~Borla, I.~Bloch, M.~Dalmonte, E.~Demler, J.-Y.~Desaules, A.~Fedorov, N.~Goldman, L.~Homeier, A.~Hudomal, C.~McCabe, I.~P.~McCulloch, K.~Jansen, V.~Kasper, M.~Kebri\v{c}, J.~Knolle, H.~Lang, S.~Linsel, J.~Mildenberger, S.~Moroz, J.~Osborne, J.-W.~Pan, Z.~Papi\'c, L.~Pollet, P.~Popov, U.~Schollw\"ock, C.~Schweizer, G.-X.~Su, P.~Stornati, E.~Tirrito, R.~Verresen, Z.-S.~Yuan, H.~Zhao, E.~Zohar for fruitful discussions and collaborations on topics related to this review.

\bibliography{biblio}

\end{document}